\documentclass[11pt,letterpaper]{article}
\usepackage[letterpaper, margin=0.75in]{geometry}

\usepackage{graphicx,color}
\usepackage[cp850]{inputenc}
\usepackage[T1]{fontenc}
\usepackage{rotating}
\usepackage[f]{esvect}
\usepackage{amsmath, amsthm, amssymb}
\usepackage{rotating}
\usepackage{lscape}
\usepackage{mathrsfs}
\usepackage{subcaption}
\usepackage{mathtools}
\usepackage[toc,page]{appendix}
\usepackage{hyphenat}
\usepackage{mathptmx}
\usepackage[scaled=.65]{helvet}
\usepackage{courier}
\usepackage{rotating}
\usepackage[right,displaymath, mathlines]{lineno}
\graphicspath{{Figures/}}

\newcommand{\ee}{{\rm e}}
\newcommand{\dd}{{\rm d}}

\usepackage[normalem]{ulem}

\DeclarePairedDelimiter\ceil{\lceil}{\rceil}
\DeclarePairedDelimiter\floor{\lfloor}{\rfloor}

\def\tablenotes{\bgroup\parfillskip=0pt plus 1fil
\leftskip=0pt\relax \rightskip=0pt
\vskip2pt\footnotesize}
\def\endtablenotes{\vskip1pt\egroup}

\def\sphline{\noalign{\vskip3pt}\hline\noalign{\vskip3pt}}

\renewcommand{\epsilon}{\varepsilon}

\renewcommand{\geq}{\geqslant}
\renewcommand{\d}{\mathrm{d}}

\renewcommand{\epsilon}{\varepsilon}

\newcommand{\TimeDeriv}{\frac{\textrm{d}}{\textrm{dt}}}

\newcommand{\HypoExp}{\mathrm{HypoExp}}

\allowdisplaybreaks

\usepackage[square,numbers,sort&compress]{natbib}
\usepackage{comment}
\usepackage{graphicx}
\usepackage{overpic} 
 \graphicspath{ {./Figures/} }
 
 \usepackage{setspace}
 \usepackage[right,mathlines]{lineno}
\let\oldequation\equation
\let\oldendequation\endequation

\renewenvironment{equation}
  {\linenomathNonumbers\oldequation}
  {\oldendequation\endlinenomath}
  
\let\oldalign\align
\let\oldendalign\endalign

\renewenvironment{align}
  {\linenomathNonumbers\oldalign}
  {\oldendalign\endlinenomath}

\renewenvironment{align*}
  {\linenomathNonumbers\oldalign\notag}
  {\notag \oldendalign \endlinenomath}

\title{An efficient and flexible generalization of transit compartment models for delayed biological processes}

 \author{Giulia Belluccini\textsuperscript{1},
 Christiaan H. van Dorp\textsuperscript{2},
 Linjie Huang\textsuperscript{1}, \\
 Jodie A. Cochrane\textsuperscript{3},
 Sara Hamis\textsuperscript{3},
 Tyler Cassidy\textsuperscript{4,*} }
\date{\today}

\begin{document}

 \maketitle

\small{ 
\textsuperscript{1}  School of Mathematics, University of Leeds, Leeds, LS2 9JT, United Kingdom. \\
\textsuperscript{2}
Department of Internal Medicine, Radboud Community for Infectious Diseases, Radboud University Medical Center, Nij\-me\-gen, Netherlands \\
\textsuperscript{3} Department of Information Technology, Uppsala University, Uppsala, Sweden. \\
\textsuperscript{4} Faculty of Pharmaceutical Sciences, University of British Columbia, Vancouver, V6T 1Z3, Canada.} \\
*\textbf{Correspondence:} tyler.cassidy@ubc.ca 
 
 \maketitle
 
 \section*{Abstract}
 
Transit compartment models are widely used in pharmacokinetics/pharmacodynamics to represent delayed biological processes such as absorption, maturation, cellular proliferation, and treatment-induced toxicity. This transit compartment formulation imposes an artificial relationship between the mean and variance of the delay distribution, as the number of transit compartment must be integer-valued number of transit compartments. Consequently, estimating the number of compartments from experimental data leads to mixed-integer optimization problems that artificially constrain model dynamics and bias parameter inference. However, these transit compartment models are mathematically equivalent to distributed delay differential equations with Erlang delay kernels. We leverage a hypoexponential generalization of these distributed delay differential equations that preserves the familiar ordinary differential equation structure while allowing the mean and variance of the delay distribution to be estimated directly as real-valued parameters. This approach does not increase the number of inferred parameters and can be immediately implemented as a straightforward modification of existing transit compartment models. We demonstrate the approach using models of viral eclipse phases, lymphocyte proliferation, and chemotherapy-induced neutropenia. Across these examples, the hypoexponential formulation provides comparable or improved fits relative to Erlang transit models and substantially reduces computational cost by avoiding grid search over compartment numbers. This framework provides a practical approach to estimate biologically meaningful delay distributions in quantitative systems pharmacology (QSP) models.

 \clearpage
 
\section{Introduction}

Time delays, such as the eclipse phase in infection dynamics \citep{Ribeiro2010,Liao2020,Holder2011}, maturation delays in hematopoiesis \citep{Rubinow1975,Mackey1977,VonSchulthess1982,Cassidy2018a}, the distinct stages underlying  the cell-cycle \citep{Cassidy2018,Yates2017,Mackey1978}, and the duration of the infectious period in epidemiology \citep{Champredon2018,Wearing2005,Scarabel2025} are ubiquitous throughout physiology and pharmacology. These delays, which are often modelled using chains of transit compartments \citep{Koch2015,Krzyzanski2011,Hu2018}, naturally lead to delay differential equations (DDEs), where the dynamics at time $t$ explicitly depend on the solution in the past. It is increasingly understood that heterogeneity in the duration of these delayed processes can be an important factor in driving biological dynamics \citep{Cassidy2024,Cassidy2018a,Koch2015,Craig2016}. This heterogeneity is particularly important in pharmacokinetics and pharmacodynamics, where variability in absorption, maturation, or cellular lifespan directly impacts observed drug responses \citep{Bauer2025,Koch2014,Hof2021}. Consequently, distributed DDEs, which account for this heterogeneity, are increasingly used throughout pharmacology \citep{Hu2018,deSouza2017,Krzyzanski2018}. Distributed DDEs are generically given by 
\begin{align}\label{Eq:GenericDistributedDDE}
    \TimeDeriv x(t) = F\left( x(t), \int_0^{\infty} x(t-s) K(s) \d s \right),
\end{align}
 where the delay kernel $K(s)$ is a probability density function (PDF) that weighs the contribution of past states and captures heterogeneity in the delayed biological process. In practice, this kernel is often chosen to be the PDF of an Erlang distribution with $K(s) = g_a^j(s)$ where 
\begin{align*}
    g_a^j(s) = \frac{a^j}{\Gamma(j)} s^{j-1}e^{-as} \hspace{3pt} \textrm{for} \hspace{2pt}  a > 0 \hspace{2pt} \textrm{and} \hspace{2pt}  j \in \mathbb{N}.
\end{align*} 

The linear chain technique \citep{Vogel1961,MacDonald1978} establishes the equivalence between these Erlang distributed DDEs with transit compartment ordinary differential equations (ODEs)  \citep{Cassidy2018,Cassidy2022,Diekmann2017,Diekmann2020a,Hurtado2020,Hurtado2019}. In these transit compartment models, transit between successive compartments occurs at a constant and fixed rate and the concatenation of these transit compartments replaces the delayed term \citep{Cassidy2020a,Diekmann2020,Hu2018,Krzyzanski2011,Krzyzanski2019,Hof2021}. The resulting ODE formulation is commonly used throughout pharmacology without necessarily considering the equivalent distributed DDE. For example, the Friberg model of chemotherapy-induced neutropenia \citep{Friberg2002}, which is the gold-standard \citep{Craig2017} model to understand hematological toxicities during chemotherapy, is an extremely well-studied transit compartment model that is equivalent to an Erlang distributed DDE \citep{deSouza2017}. 
 
Using the ODE formulation of the transit compartment models, rather than the equivalent distributed DDE, is extremely computationally efficient, as standard ODE solvers can be used during parameter inference and model simulation. However, the ODE formulation imposes an implicit relationship between the mean and variance of the delayed process being modelled. Specifically, the mean delay $\tau = j/a$ and variance $\sigma^2 = j/a^2$ determine the shape and scale parameters of the Erlang distribution, $j$ and $a$. As shape parameter $j$ (or number of compartments) must be an integer, we must have 
\begin{align*}
 \tau^2/\sigma^2  \in \mathbb{N}.
\end{align*}
However, there is no biological justification for this relationship between the mean and variance of the delayed process. Furthermore, the constraint $j \in \mathbb{N}$ leads to mixed-integer optimization problems when performing parameter inference. These mixed-optimization problems do not easily permit gradient-based inference methods and are not implemented in existing scientific software such as Stan or Monolix \citep{Carpenter2017,Monolix2020}. Consequently, modellers often estimate model parameters for distinct, fixed values of $j$, and compare model fits via information criterion or other measurements of goodness-of-fit \citep{Beauchemin2017,Vehtari2017}, which results in computational inefficiencies to determine the number of compartments $j$ \citep{Hu2018}. However, the number of compartments does not have an immediate biological interpretation \citep{Krzyzanski2018}; rather, the mean and variance of the delayed process are of more biological interest.

Consequently, there has been increased interest in developing numerical and mathematical techniques to permit more general delay distributions in Eq.~\eqref{Eq:GenericDistributedDDE} while maintaining the numerical tractability of ODE models. For example, \citet{Hurtado2019} established a framework to derive ODE representations of distributed DDEs with phase-type delay distributions. These phase-type delay distributions correspond to hitting times in continuous Markov processes and the framework developed by \citet{Hurtado2019} represents a specific case of Theorem 4.1 of \citet{Diekmann2017}. These phase-type distributions are dense in the space of positive distributions, but require a significant increase in the number of free parameters to be inferred from data. Alternatively, recent work now allows for direct implementation of DDE solvers in Monolix or NONMEM through the implementation of quadrature methods and discrete DDE solvers \citep{Bauer2025}. In this work, we eliminate the need for integer-valued transit structures and enable direct estimation of biologically meaningful delay distributions, \textit{without} increasing model complexity or requiring the implementation of DDE-specific numerical methods. 

Specifically, rather than using the common Erlang distributed DDE framework, we consider hypoexponentially distributed DDEs. Essentially, these hypoexponential distributed DDEs generalize the common Erlang distributed DDEs and are amenable to straightforward extension of the linear chain technique that yields equivalent ODE reductions with the familiar transit compartment interpretation \citep{Cassidy2022,Hurtado2019,Hurtado2020}. These equivalent ODE systems allow the transit rates between consecutive compartments to vary \citep{Cassidy2022}, and we have previously shown that these transit rates can be chosen to capture the mean and variance of the underlying gamma distributed delayed process using the \textit{same number} of parameters as the common Erlang approximation, although the shape parameter is no longer constrained to be integer-valued. Consequently, the hypoexponential approach removes the implicit relationship between $\tau$ and $\sigma^2$ while retaining the ODE reducibility that allows for efficient parameter inference and avoids the computational difficulties posed by the mixed-optimization problem for Erlang distributed DDEs \citep{Cassidy2022}. We extend the previously developed transit compartment models of \citet{Beauchemin2017,Belluccini2022,Friberg2002} to include the more general hypoexponential delay distribution and illustrate the benefits of the hypoexponential framework by fitting these more general models. These examples show how using a hypoexponential, instead of Erlang, distribution can facilitate improved model fits while decreasing computational cost. While we focus on these examples, this method immediately extends to any transit compartment model and eliminates the need to specify an integer number of transit compartments but preserves the computationally efficient ODE representation.

\section{Methods} 

\subsection{Theoretical considerations related to distributed DDEs}

While we focus on models that were first presented using a series of transit compartments to represent a delayed biological process, these models also admit equivalent formulations as distributed DDEs. Here, we give background results focused on the relationship between these transit compartment models and distributed DDEs. However, a more mathematical treatment, including numerical methods, linearisation, and the corresponding phase spaces, of these infinite delay DDEs are available \citep{Diekmann2012,Hale1993,Cassidy2022,Hino1991,Diekmann2020c}.

In general, distributed DDEs are given in Eq.~\eqref{Eq:GenericDistributedDDE} where the PDF $K(s)$ is the Radon-Nikodym derivative of the probability measure $\mathbb{P}_Y$ with respect to the Lebesgue measure.  Here, we explicitly consider Erlang and hypoexponential distributed delays. Hypoexponential distributions arise from concatenating exponential distributed random variables in the same manner as the familiar Erlang distributions. However, unlike Erlang distributions, where each exponential distribution has the same rate, hypoexponential distributions allow for the constituent exponential distributions to have different rates. We denote a hypoexponential distribution with rates $a_1,a_2,...,a_j$ by $\HypoExp(a_1, a_2,\dots, a_j)$. Further, we note that if $ a_1 = a_2=...=a_j = a,$ then $\HypoExp(a, a,\dots, a)$ is precisely the Erlang distribution with shape parameter $j$ and rate parameter $a$. Thus, hypoexponential distributions generalize Erlang distributions. 
\subsubsection{The linear chain technique}

The linear chain technique has been known since at least the work of \citet{Vogel1961} and was popularised by \citet{MacDonald1978}. In general, the linear chain technique replaces the semi-infinite convolution integral that encodes the delay in Eq.~\eqref{Eq:GenericDistributedDDE} by the solution of an auxiliary system of linear ODEs. While the linear chain technique traditionally derives an system of ODEs that is mathematically equivalent to a distributed DDE, this equivalence also applies in the opposite direction. Specifically, for a given transit compartment model, it is possible to derive the equivalent distributed DDE using the same approach, which may simplify analytical study of the model \citep{deSouza2017,Cassidy2020a}. 

For an Erlang distributed DDE with PDF $K(s) = g_a^j(s)$ and $x(t-s)$ representing an inflow rate in units of population/time, we define the auxiliary variables
\begin{align*}
    A_n (t) = \int_0^{\infty} \frac{x(t-s)}{a} g_a^n(s)\d s,
\end{align*}
 for $n = 1,2,...,j$, and the division by $a$ ensures that $A_n$ has the units of the population of interest. We note that $aA_j(t)$ is precisely the convolution integral in the Erlang distributed DDE, Eq.~\eqref{Eq:GenericDistributedDDE}. Then, a straightforward application of Leibniz's rule shows that 
 \begin{align*}
 \TimeDeriv A_1(t) & =  x(t)- aA_1(t). \\
 \TimeDeriv A_n(t) & = a\left[A_{n-1}(t)-A_n(t)\right] \quad \textrm{for} \quad n = 2,3,...,j, 
 \end{align*}
In either the ODE or DDE case, modellers need to specify two free parameters, the rate $a$ and shape $j$ of the Erlang distribution. As mentioned, these two parameters determine the mean ($\tau = j/a$) and variance ($\sigma^2 = j/a^2$) of the distribution with the constraint that $j = \tau^2/ \sigma^2 \in \mathbb{N}.$ Thus, when using Erlang distributions or the equivalent transit compartment formulation, it is common to prescribe the number of compartments $j$ and the rate $a$, rather than the mean and variance.

Now, consider the more general case where the delay follows a hypoexponential distribution. The corresponding DDE is given by
\begin{align*} 
    \TimeDeriv x(t) = F \left( x(t),\int_0^{\infty} x(t-s) \d \mathbb{P}_{\mathcal{Y}} \right),
\end{align*}
where $\mathcal{Y} \sim \HypoExp(a_1, a_2, ...,a_j)$. In general, modellers must prescribe the $n$ rates of the hypoexponential distribution $a_1, a_2, ..., a_j$. Then, we define auxiliary variables $B_i(t)$ by 
\begin{align*}
 B_i (t) = \int_0^{\infty} \frac{x(t-s)}{a_i}  \d \mathbb{P}_{\mathcal{Y}_i},  
\end{align*}
where $ \mathcal{Y}_i \sim \HypoExp(a_1, a_2, ...,a_i)$ is the hypoexponential distribution defined by the first $i$ rates  $a_1, a_2, ..., a_i$. These auxiliary variables satisfy the equivalent system of ODEs \citep{Hurtado2019,Cassidy2022,Cassidy2020a}
\begin{align*} 
        \TimeDeriv B_1(t) & = x(t) - a_1B_1(t) \\
        \TimeDeriv B_{i}(t) & = a_{i-1}B_{i-1}(t) -a_i B_i(t) \quad \textrm{for} \quad i = 2, 3, ...,j. 
\end{align*}

Now, consider a hypoexponential distribution with mean delay $\tau$ and variance $\sigma^2$ with no \textit{a priori} relationship between these two moments. Then, \citet{Cassidy2022} showed how to choose the rates $ a_i $ to ensure that the resulting hypoexponential distribution has mean $\tau$ and variance $\sigma^2$ so that the distribution is a continuous function of the rates $a_i$ via the so-called \textit{smoothed approximation}. 

Specifically, set $n = \max\left[2,  \ceil{\tau^2/\sigma^2} \right]$, take $ j  = \tau^2/\sigma^2 \in \mathbb{R}$, and let $\{j \} = j - \floor{j}$ be the fractional part of $j$. For 
\begin{align}\label{Eq:HypoexponentialParametrisation}
    \lambda = j/\tau, \quad \frac{1}{\nu} = \frac{\tau}{2j} \left( 1 + \{j \} +  \sqrt{1-\{ j \}^2} \right) \quad \textrm{and} \quad 
     \frac{1}{\mu} = \frac{\tau}{2j} \left( 1 + \{j \} -  \sqrt{1-\{ j \}^2} \right) ,
\end{align}
with $ \mu = \nu = \lambda$ if $j \in \mathbb{N},$ Theorem 3.2 of \citet{Cassidy2022} shows that $\HypoExp(\lambda,\lambda,...,\lambda,\mu,\nu)$ has mean $\tau$ and variance $\sigma^2.$ Thus, the first two moments of the hypoexponential distribution completely parameterize the $n$ unknown rates to match the given mean and variance. Modellers do not need to prescribe all $j$ rates of the hypoexponential distribution, but rather only prescribe the mean $\tau$ and variance $\sigma^2$. This ensures that the hypoexponential and Erlang distributions therefore have the same number of free parameters, and there is no increase in model complexity between Erlang and hypoexponential distributed DDEs. The choice of the rates $\{ \lambda, \mu, \nu \}$ are continuous functions of $j$ except in the case $j \in \mathbb{N}$, which facilitates parameter estimation.     

%

\subsection{Experimental data} \label{Sec:ExperimentalData}

As mentioned, we estimate model parameters using data from previously modelled biological systems. Here, we discuss the data used in the remainder of our study.

\subsubsection{Duration of the eclipse phase in SHIV infection}

\citet{Beauchemin2017} considered the duration of the eclipse and virus producing phases of simian CD4+ T-cells (HSC-F cells) infected by the SHIV-KS661 strain of simian-human immunodeficiency virus. There, the eclipse phase was defined as the period between successful infection of a host cell and the release of the first infectious virion by the newly infected cell. Similarly, the virus producing phase corresponds to the interval between the release of the first and last virus particle for a given infected cell. \citet{Beauchemin2017} measured the fraction of infected and dead cells as well as the viral load concentration for five serial dilutions of initial multiplicities of infection (MOI), beginning at $4.2\times 10^6$ TCID/mL, at eight different time points, leading to 120 total measurements. We digitized the data from \citet{Beauchemin2017} using the \texttt{digitize2} function in MATLAB \citep{Matlab2017}. We validated this digitization by simulating the \citet{Beauchemin2017} model using the reported best-fit parameter estimates from \citet{Beauchemin2017} and obtained the same sum of squares residuals (SSR) reported by \citet{Beauchemin2017} for these parameter values. 

When considering samples taken at time $t_i$ from the $j$th dilution, we denote the total viral load concentration by $V_{RNA,j,i}^{data}$, the concentration of infected cells by $ I_{j,i}^{data}$, and the concentration of dead or no-longer infectious cells by $D_{j,i}^{data}$. 

\subsubsection{Lymphocyte generation times}

We consider experimental data from lymphocyte labeling experiments \cite{lyons1994determination} that quantify T-cell proliferation \textit{in vivo}. Here, the level of intracellular dilution of the label reflects the number of generations $G$ of a cell. \citet{hogan2013clonally} considered the dynamics of OT-I and F5 T-cells, which are characterised by different rates of lymphopenia-induced proliferation. Lymphopenic mice were injected with CFSE-labelled cells. Then, on days 3, 4, 5, 6, 7, 10, 12 and 18 following infusion of the labelled cells, 3-7 mice were sacrificed and analysed by flow cytometry to quantify levels of CFSE dilution, and thus the generation of the T-cells \citep{hogan2013clonally}. All cells in generation 5 or higher were considered as a single generation, denoted $5+$~\citep{Belluccini2022}. Following \citet{Belluccini2022}, we only consider the experimental data during exponential growth of these cell-lines and so only consider the OT-I data until day 7.

For both cell lines, denote the number of cells in the $G$-th generation at time $t_i$ by $M_G^{data}(t_i)$. 

\subsubsection{Chemotherapy-induced neutropenia}

\citet{Craig2016} developed a mathematical model of circulating neutrophil dynamics following chemotherapy. The model is a system of state-dependent DDEs that was fit to circulating neutrophil concentrations from a clinical trial of the chemotherapeutic Zalypsis. As a validation step, \citet{Craig2016} accurately predicted the neutrophil response to CHOP14 chemotherapy. 

Here, we generated synthetic data representing neutrophil concentrations by simulating the \citet{Craig2016} model of neutrophil dynamics following CHOP14 chemotherapy. Specifically, we simulated three 21-day cycles of CHOP14 chemotherapy and sampled the simulated circulating neutrophil concentration on days 0, 6, and every 3 days until day 69. We added random noise to each of these 21 measurements. As the \citet{Craig2016} model was validated by predicting the neutrophil response to anti-cancer chemotherapy, this synthetic data is a reasonable proxy for clinical data, and this sampling frequency is similar to the clinical data considered by \citet{Friberg2002}. Consequently, we consider the simulated data 
\begin{align*}
    N_{i}^{data} = N_{sim}(t_i) + 0.1 e_i,
\end{align*}
where $e_i \sim \mathcal{N}(0,1) $, $t_i$ is the sample time, and $N_{sim}(t_i)$ is the simulated neutrophil concentration at time $t_i$ using the \citet{Craig2016} model. The synthetic data is strictly positive even after the addition of the noise.

\subsection{Existing mathematical models}

In this work, we consider three distinct mathematical models for delayed processes in biology. These models were first presented as transit compartmental ODEs \citep{Beauchemin2017, Belluccini2022, Friberg2002}. We now derive the equivalent distributed DDE representation for each of the models.  

\subsubsection{Eclipse phase in viral dynamics}

\citet{Beauchemin2017} adapted the standard viral dynamics model \citep{Perelson2002,Perelson1996} to study the duration of the eclipse and infectious phases of \textit{in vitro} SHIV viral dynamics. The \citet{Beauchemin2017} viral dynamics model tracks the concentration of uninfected target cells, denoted by $T(t)$, infectious free virus, $V_{TCID}(t)$, total viral RNA $V_{RNA}(t)$, and infected cells $I(t)$. When modelling these \textit{in vitro} assays, we do not model any proliferation or death of uninfected target cells \citep{Beauchemin2017}, so the dynamics of target cells are driven only by infection. Infection occurs at a rate $\beta$ following interaction between free virus and target cells. Following an Erlang-distributed eclipse phase with mean duration $\tau_E$ and shape parameter $n_E$, infectious cells produce infectious virus at a constant rate $p_{TCID}$ and viral RNA at a constant rate $p_{RNA}$. \citet{Beauchemin2017} also assume that the lifespan of infectious cells is Erlang-distributed with mean duration $\tau_I$ and shape parameter $n_I,$ and explicitly track the concentration of non-productive, infected cells $D(t)$. The infectious and total virus are cleared at constant rates $c_{TCID}$ and $c_{RNA}$, respectively. We give a graphical representation of the \citet{Beauchemin2017} model in Fig~\ref{Fig:BeaucheminFigure}A. Using the linear chain technique to transform the transit compartment~\citet{Beauchemin2017} viral dynamics model  to the corresponding Erlang distributed DDE, the model of \textit{in vitro} SHIV infection is
 \begin{equation} \label{Eq:BeaucheminDDE}
 \left.
 \begin{aligned} 
     \TimeDeriv T(t) & = -\beta  T(t) V_{TCID}(t)\\
     \TimeDeriv I(t) & = \int_{-\infty}^{t} \beta V_{TCID}(s) T(s)g_{a_E}^{n_E}(t-s) \d s  -  \int_0^{\infty} \left[ \int_{-\infty}^{t-\sigma} \beta V_{TCID}(s) T(s)g_{a_E}^{n_E}(t-\sigma-s) \d s \right] g_{a_I}^{n_I}( \sigma)  \d \sigma \\
     \TimeDeriv D(t) & = \int_0^{\infty} \int_{-\infty}^{t-\sigma} \beta V_{TCID}(s) T(s)g_{a_E}^{n_E}(t-\sigma-s)  g_{a_I}^{n_I}( \sigma) \d s \d \sigma \\
     \TimeDeriv V_{TCID}(t) & = p_{TCID}I(t) - c_{TCID}V_{TCID}(t) \\ 
     \TimeDeriv V_{RNA}(t) & = p_{RNA}I(t) - c_{RNA}V_{RNA}(t). 
 \end{aligned} 
 \right \}
 \end{equation}
The differential equation for target cells corresponds solely to infection with rate $\beta T(t) V_{TCID}(t)$ and is unchanged from \citet{Beauchemin2017}. Now, the first delay term in Eq.~\eqref{Eq:BeaucheminDDE} corresponds to the rate that cells complete the eclipse phase at time $t$. These cells were infected at time $t-s$ where $s$ is the duration of the eclipse phase. This duration follows an Erlang distribution with shape parameter $n_E$ and rate parameter $a_E = n_E/\tau_E.$ 

Then, the second integral term gives the rate at which infectious cells are cleared at time $t$ after an infectious period of duration $\sigma$. Here, the infectious phase follows an Erlang distribution with shape parameter $n_I$ and rate parameter $a_I = n_I/\tau_I$. These infectious cells completed the eclipse phase, and thus entered the productive infection period, at time $t-\sigma$. Cells that completed the eclipse phase at time $t-\sigma$ were infected at time $t-\sigma-s$, where $s$ is the duration of the eclipse phase. A formal derivation of the distributed DDE for $I(t)$ can be obtained from the equivalent renewal equation formulation \citep{Diekmann2012a,Cassidy2022,Cassidy2020a}.

\sloppy
Replacing these Erlang distributed delays for the eclipse and infectious periods by hypoexponential distributions $\HypoExp_E (a_{E,1}, a_{E,2},...,a_{E,n_E}) $ and $\HypoExp_I(a_{I,1}, a_{I,2},...,a_{I,n_I})$ allows for the following ODE representation of Eq.~\eqref{Eq:BeaucheminDDE}
 \begin{equation} \label{Eq:BeaucheminHypoexponential}
 \left.
 \begin{aligned} 
     \TimeDeriv T(t) & = -\beta  V_{TCID}(t)T(t) \\
     \TimeDeriv E_1(t) & = \beta  V_{TCID}(t)T(t) - a_{E,1}E_1(t) \\
     \TimeDeriv E_i(t) & = a_{E,i-1}E_{i-1}(t) - a_{E,i}E_i(t), \quad \textrm{for} \quad i = 2,3,...,n_E \\
     \TimeDeriv I_1(t) & =   a_{E,n_E}E_{n_E}(t) - a_{I,1}I_1(t) \\
     \TimeDeriv I_i(t) & = a_{I,i-1}I_{i-1}(t) - a_{I,i}I_i(t), \quad \textrm{for} \quad i = 2,3,...,n_I \\ 
     \TimeDeriv D(t) & =  a_{I,n_I}I_{n_I}(t) \\ 
     \TimeDeriv V_{TCID}(t) & = p_{TCID}I(t) - c_{TCID}V(t) \\ 
     \TimeDeriv V_{RNA}(t) & = p_{RNA}I(t) - c_{RNA}V(t),  
 \end{aligned} 
 \right \}
 \end{equation}
where 
\begin{align*}
    I(t) = \displaystyle \sum_{i=1}^{n_I} I_i(t),
\end{align*}
 and the initial conditions correspond to zero cells in either the eclipse or infected phases. 
It is simple to note that, if $a_{E,i} = n_E/\tau_E$ and $a_{I,i} = n_I/\tau_I$ for all $i$, then Eq.~\eqref{Eq:BeaucheminHypoexponential} is precisely the transit compartment model utilised by \citet{Beauchemin2017}.

\subsubsection{Generation times in lymphocyte dynamics} 
 
\citet{Belluccini2022} developed a multi-stage mathematical model to quantify the birth and death rates of activated lymphocyte populations. Let $M_G(t)$ denote the number of cells in generation $G$ at time $t$.  In this model, the duration of the $G$-th generation is Erlang distributed with shape parameter $N_G$ and scale parameter $\lambda_G$ while cells in generation $G$ may die with per capita death rate $\mu_G$. Consequently, this model assumes that, given a cell in generation $G$, its time to division follows an Erlang distribution while the time to death follows an exponential distribution with rate $\mu_G$.  Proliferating cells in generation $G$ give rise to two daughter cells in generation $G+1$. The model dynamics are depicted in Fig.~\ref{Fig:BellucciniFig}A. The corresponding Erlang distributed DDE is given by  
\begin{equation}\label{Eq:LymphocyteGenerationsDDE}
\left.
    \begin{aligned} 
        \TimeDeriv M_0(t) & = - \int_{-\infty}^{t} \lambda_0 M_0(s) \ee^{-\mu_0 (t-s)} g^{N_0}_{\lambda_0} (t-s) \dd s - \mu_0 M_0(t), \\
        \TimeDeriv M_G(t) & =   2 \int_{-\infty}^{t}   \lambda_0 M_{G-1}(s) e^{-\mu_{G-1} (t-s)} g^{N_{G-1}}_{\lambda_{G-1}} (t-s) \d s- \mu_G M_G(t),  \\
        & {}  - \int_{0}^{\infty}  \left[  2 \int_{-\infty}^{t-\sigma}  \lambda_{G-1} M_{G-1}(s) e^{-\mu_{G-1}(t-\sigma - s) } g^{N_{G-1}}_{\lambda_{G-1}} (t-\sigma - s) \d s \right] e^{-\mu_{G} \sigma } g^{N_{G}}_{\lambda_{G}} ( \sigma ) \d \sigma \quad \textrm{for}  \quad G = 1,2,\dots ,
    \end{aligned}
    \right \}
\end{equation}
where the first term of the latter equation accounts for the arrival in generation $G$ of two cells generated by the division of a cell in generation $G-1$, and the final two terms reflect a cell lost in generation $G$ due to death or division, respectively. The final term captures the distributed residence time of cells in generation $G$ who entered this generation at time $t-\sigma$ in the past. This expression can be obtained from the chain-like structure of generation dynamics \citep{Cassidy2020a}. Replacing the Erlang distributed delay for the time to division of a cell in generation G by the hypoexponential distribution $\HypoExp_E (a_{G,1}, a_{G,2},...,a_{G,N_G})$, we obtain the following ODE representation of Eq.~\eqref{Eq:LymphocyteGenerationsDDE} 
\begin{equation} \label{Eq:LymphocyteGenerationsHypo}
\left.
    \begin{aligned}
        \TimeDeriv T_{0,1}(t) & = -(a_{0,1} + \mu_0) T_{0,1}(t), \\
        \TimeDeriv T_{G,1}(t) & = 2 a_{G-1, N_{G-1}} T_{G-1, N_{G-1}}(t) - (a_{G,1} + \mu_G) T_{G,1}(t), \quad \text{for} \quad G = 1,2,\dots , \\
        \TimeDeriv T_{G,i}(t) & = a_{G,i-1} T_{G,i-1}(t) - (a_{G,i} + \mu_G) T_{G,i}(t), \quad \text{for} \quad G = 1,2,\dots, \quad \textrm{and}  \quad i = 2,3,\dots,N_G,
    \end{aligned}
\right \}
\end{equation}
where $T_{G,i}(t)$ denotes the number of cells in stage $i$ of generation $G$ at time $t$,
\begin{equation}\label{Eq:cells_in_generat}
    M_{G}(t) = \sum_{i=1}^{N_G} T_{G,i}(t).
\end{equation}
The initial conditions for the transit compartments correspond to zero cells in any of the compartments except $T_{0,1}$.

\subsubsection{Chemotherapy induced neutropenia}

\begin{figure}[ht]
\centering
\includegraphics[scale=0.99, trim= 20 10 20 10,clip]{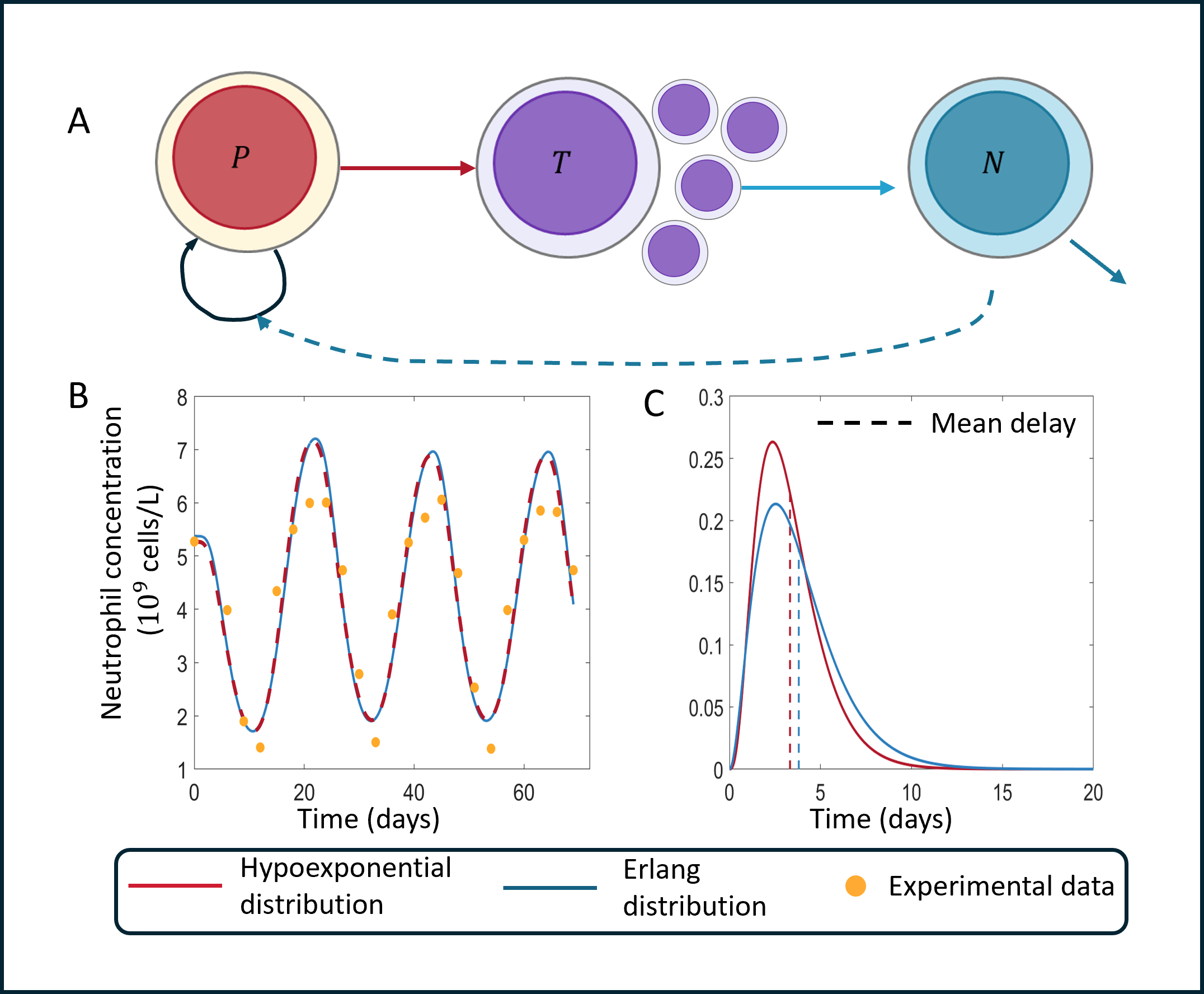}   
\caption{  \textbf{A)} Schematic representation of the \citet{Friberg2002} model. \textbf{B)} Model fits to synthetic neutropenia data generated using the \citet{Craig2016} model. The synthetic data is plotted as filled circles and the model predictions are continuous curves. The red curve is the model fit using the hypoexponential distribution and the blue curve is the model fit using the Erlang distribution. \textbf{C)} Comparison of the distribution PDFs for the duration of the neutrophil maturation period time. The dashed vertical lines give the mean of each distribution.  }
\label{Fig:FribergFigure}
\end{figure}

Neutrophils are the most common white blood cells in humans and an integral part of the immune system. Neutropenia is a common dose-limiting toxicity of traditional anti-cancer chemotherapies \citep{Roskos2006,deSouza2017,Dieras2019}. \citet{Friberg2002} developed a semi-mechanistic transit compartmental model of neutrophil production. This transit compartmental model has become the gold-standard framework for modelling hematopoietic toxicity and has been used throughout the pharmaceutical sciences \citep{Craig2017,Soto2020,Quartino2014}. The \citet{Friberg2002} model is a system of differential equations that model the dynamics of proliferating neutrophil precursors ($P(t)$), $M$ compartments representing maturation stages in the bone marrow ($T_{i}(t)$), and a final circulating compartment that can be directly compared against measured data ($N(t)$).  \citet{Friberg2002} considered $M =3$ corresponding to three maturation compartments and \citet{deSouza2017} showed how to convert the \citet{Friberg2002} model to the equivalent system of Erlang distributed DDEs with shape parameter $M$ and maturation rate $a$  
 \begin{equation} \label{Eq:FribergDDE}
 \left.
 \begin{aligned} 
     \TimeDeriv P(t) & =  \left[ k_p(1-E_{drug}(t))\left( \frac{N_0}{N(t)} \right)^{\gamma} -k_{tr} \right] P(t)\\
     \TimeDeriv N(t) & =  \int_0^{\infty} k_{tr} P(t-s) g_a^M(s) \d s - k_{circ} N(t).
 \end{aligned} 
 \right \}
 \end{equation}
In Eq.~\eqref{Eq:FribergDDE}, $k_p$ is the baseline proliferation rate, the ratio $N_0/N(t)$ represents the feedback from circulation on proliferating neutrophil precursors, $\gamma$, controls the strength of this feedback, and $k_{tr}$ models transit from the proliferating population into the maturation stage. Here, we follow \citet{deSouza2017} and distinguish between the rate at which cells enter maturation, $k_{tr}$, and the maturation rate, $a$. Furthermore, for the model to be in equilibrium when $N(t) = N_0$, which we take as the initial condition, we must enforce $k_{tr} = k_p$ \citep{Quartino2014,deSouza2017}. Finally, circulating neutrophils are cleared at a constant rate $k_{circ}.$ We give a schematic diagram of the \citet{Friberg2002} model for neutrophil dynamics in Fig~\ref{Fig:FribergFigure}A.

Then, replacing the Erlang distributed maturation delay in Eq.~\eqref{Eq:FribergDDE} by a hypoexponential distribution with $M$ compartments 
 $\HypoExp(a_1,...,a_M)$ gives the equivalent system of ODEs, where we recall that $k_{tr} = k_p,$ 
 \begin{equation} \label{Eq:FribergHypoexponential}
 \left.
 \begin{aligned} 
     \TimeDeriv P(t) & =  \left[ (1-E_{drug}(t))\left( \frac{N_0}{N(t)} \right)^{\gamma}  -1 \right] k_pP(t) \\
     \TimeDeriv T_1 & = k_p P(t) - a_1T_1(t) \\
     \TimeDeriv T_{i} & = a_{i-1} T_{i-1}(t) - a_iT_i(t) \quad \textrm{for} \quad i = 2,3,...,M. \\
     \TimeDeriv N(t) & =   a_M T_{M}(t) - k_{circ} N(t).
 \end{aligned} 
 \right \}
 \end{equation} 

Finally, we couple the semi-mechanistic model Eq.~\eqref{Eq:FribergDDE} with the four-compartment population pharmacokinetic (PK) model for the chemotherapeutic Zalypsis to model chemotherapy induced neutropenia \citep{Perez-Ruixo2012,Craig2016}. The PK dynamics of Zalypsis are given by 
 \begin{equation*}
 \begin{aligned}
     \TimeDeriv C_p(t) & = I_c(t) + k_{fp} C_f(t) + k_{sl1p} C_{sl1}(t) - (k_{pf} + k_{psl1} + k_{cl} ) C_p(t) \\
     \TimeDeriv C_f(t) & = k_{pf}C_p(t) + k_{sl2f}C_{sl2}(t) - (k_{fp} + k_{fsl2})C_f(t)    \\
     \TimeDeriv C_{sl1}(t) & = k_{psl1}C_p(t) - k_{sl1p} C_{sl1}(t) \\
     \TimeDeriv C_{sl2}(t) & = k_{psl2} C_p(t) - k_{sl2f} C_{sl2}(t),
 \end{aligned}
 \end{equation*}
where $C_p(t)$ represents the concentration in the central compartment, $C_f(t)$ is the concentration in the fast-exchanging tissues, and $C_{sl1}(t)$ and $C_{sl2}(t)$ are the concentrations in the slow-exchange tissues. The rates $k_{ij}$ give the transit rates from compartment $i$ to compartment $j$ and $I_c(t)$ models the subcutaneous administration of Zalypsis. We model the effect of chemotherapy on blocking effective proliferation of neutrophil precursors using an Emax model given by 
\begin{align*}
    1-E_{drug}(t) = 1- \frac{E_{max}C_p(t)}{EC_{50} + C_p(t)},
\end{align*}
where $E_{max}$ is the maximal inhibition of proliferation of neutrophil precursors and $EC_{50}$ is the half-effect concentration. 

\subsection{Parameter estimation and model selection}
The mathematical models considered in this work have a number of unknown parameters. In previous work, these parameters were estimated by fitting the model to the experimental data discussed in Section~\ref{Sec:ExperimentalData} \citep{Beauchemin2008,Belluccini2022}. To facilitate comparison between prior modelling using the Erlang approximation and the proposed hypoexponential approximations, we leverage the parameter estimation procedure for the unknown mechanistic parameters present in each model. We then compare these previous results using the Akaike Information Criterion (AIC)~\cite{anderson2004model}. As the models have the same number of free parameters and are fitting the same data, this is equivalent to comparing the models via the log-likelihoood, or via the corrected AIC to account for the relatively small amount of data.  

We also compare the clock-time for the parameter inference when using the Erlang and hypoexponential models. We performed all simulations using the same optimization settings and personal laptop. Finally, all the examples here perform parameter inference on model quantities that are defined in the equivalent distributed DDE formulation, for which we use an equivalent ODE representation. Consequently, changes in the dimension of the underlying ODE changes due to changes in the parameter $j$ do not cause discontinuities in the optimization problem, as the underlying DDE is a continuous function of the parameters describing the delay distribution. 

\subsubsection{Eclipse phase in viral dynamics}

\citet{Beauchemin2017} fix the clearance rates of infectious virus and viral RNA as $c_{TCID} = 0.93$/day and $c_{RNA} = 0.039$/day based on prior experimental and modelling work. Further, they link the production of infectious virus and total viral RNA via $p_{TCID} = 8.63 \times 10^{-6} p_{RNA}. $ Then, they estimate the infectivity rate $\beta$ and the production rate of total RNA $p_{RNA}$. Finally, they also estimate the initial infectious viral load denoted by $[\textrm{BASE}_{MOI}]$. 

Consequently, in each of the serial dilutions,  the initial concentration of infectious virus, $V_{TCID}(0)$, is given by 
\begin{align*}
    V_{TCID,i}(0) = \frac{[\textrm{BASE}_{MOI}]}{2^i} \quad \textrm{for}  \quad i = 0,1,2,3,4. 
\end{align*}
However, \citet{Beauchemin2017} note that the residual virus post-rinse was not proportional to the inocula at the two highest MOIs. We follow their implementation of the initial conditions and thus apply a proportional rinsing factor for these higher MOIs \citep{Beauchemin2017}. 

In addition to these three mechanistic parameters and the initial conditions, \citet{Beauchemin2017} estimated the number of compartments in the eclipse, $n_E$, and infectious, $n_I,$ phases by sampling integer values in $[1,100].$ They then estimated the mean duration of both the eclipse and infectious phases, $\tau_E$ and $\tau_I,$ respectively, for each pair of integers $(n_E,n_I).$ Thus, for a given parameter estimation, \citet{Beauchemin2017}  estimated five strictly positive parameters $\beta, p_{RNA},\textrm{BASE}_{MOI}, \tau_E$, and $\tau_I$, and  two integer-valued parameters $n_E$ and $n_I$.

 For a given vector of model parameters $\theta_{Eclipse} = [\beta, p_{RNA},\textrm{BASE}_{MOI}, \tau_E,\tau_I, n_E,n_I]$, the SSR is given by 
 \begin{equation}\label{Eq:BeaucheminMaximumLikelihood} 
 \left. 
\begin{aligned} 
    SSR(\theta_{Eclipse} )&  = \displaystyle \sum_{j=1}^{5} \left( \sum_{i=1}^{8} \left[ \frac{ \log_{10} ( V_{RNA}(t_{j,i},\theta_{Eclipse} )) -\log_{10}(V_{RNA,j,i}^{data}) }{s_{v} }\right]^2  \right. \\ 
    & \quad \left.  + \left[ \frac{I(t_{j,i},\theta_{Eclipse} ) -I_{j,i}^{data} }{s_{i}}\right]^2 + \left[ \frac{D(t_{j,i},\theta_{Eclipse} ) -D_{j,i}^{data} }{s_{d} }\right]^2  \right).
\end{aligned}
\right \}
\end{equation}
In Eq.~\eqref{Eq:BeaucheminMaximumLikelihood}, $V_{RNA}(t_{j,i},\theta_{Eclipse}), I(t_{j,i},\theta_{Eclipse}),$ and $D(t_{j,i},\theta_{Eclipse})$ are the simulated concentrations of total viral RNA, productively infected cells, and dead cells for the $j$th dilution at time $t_{i}$ with parameters $\theta_{Eclipse}$. \citet{Beauchemin2017} estimated the five, real valued model parameters coupled with various fixed values of $n_E$ and $n_I$ by minimizing the least-squares residual between their model prediction and the experimental data using successive applications of Levenberg-Marquardt and Nelder-Mead algorithms. The values $s_{v},s_{i}$ and $ s_{d}$ were calculated by \citet{Beauchemin2017} using the standard deviations of the pooled residuals for the 100 best fits. For consistency, we use the same values as \citet{Beauchemin2017} when evaluating Eq.~\eqref{Eq:BeaucheminMaximumLikelihood} and we follow \citet{Beauchemin2017} and consider the likelihood defined by $\exp(-SSR/2)$.

To fit the generalized \citet{Beauchemin2017} model Eq.~\eqref{Eq:BeaucheminHypoexponential}, to the experimental data, we minimized the SSR in Eq.~~\eqref{Eq:BeaucheminMaximumLikelihood}, or maximized the log-likelihood, using the \texttt{fmincon} algorithm in \citet{Matlab2017} starting from 10 initial guesses. We fit seven, real valued, model parameters given by $\theta = [\beta, p_{RNA},\textrm{BASE}_{MOI}, \tau_E,\tau_I, \sigma^2_E,\sigma^2_I]$ and used the relationships in Eq.~\eqref{Eq:HypoexponentialParametrisation} to calculate the rates $a_i$ of the hypoexponential distribution for the given pairs $(\tau_E,\sigma^2_E)$ and $(\tau_I, \sigma^2_I)$.  

\subsubsection{Generation time in lymphocyte dynamics}

\citet{Belluccini2022} fit the generational model Eq.~\eqref{Eq:LymphocyteGenerationsHypo} to the F5 and OT-I data separately. In both cases, the time to first division is larger than the time to subsequent divisions~\cite{hawkins2007model,kinjyo2015real}. Consequently, \citet{Belluccini2022} allowed the parameters determining the proliferation rate of generation 0 cells to differ from those of cells which divided at least once. Specifically, they estimated the number of stages and the birth rate of generation 0, given by $N_0$ and $\lambda_0$ separately from the number of stages and the birth rate of the remaining generations $G \ge 1$, which are denoted by $N$ and $\lambda$, respectively~\cite{Belluccini2022}. Further, lymphocytes in later generations are more likely to undergo apoptosis ~\cite{ganusov20072}. Therefore, \citet{Belluccini2022} assumed that the death rate $\mu$ increases as a function of cell generation given by  $\mu_G = \alpha G$, $G \ge 0$, where $\alpha$ is a parameter to estimate ~\cite{mazzocco2017estimates,Belluccini2022}. \citet{Belluccini2022} also estimate the initial number of cells, $C_0$, so the model parameters $\theta  = [N_0, \lambda_0, N, \lambda, \alpha, C_0]$ are inferred by minimizing
\begin{equation} \label{Eq:LymphocyteMaximumLikelihood}
        SSR(\theta ) = \sqrt{\sum_{G=0}^{5+} \sum_{t \in \mathscr{T}} \left[ \frac{\log_{10} (M_G(t,\theta)) - \log_{10} (M_G^{data}(t_i))}{\sigma_G^{data}(t_i)} \right]^2},
\end{equation}
where $\mathscr{T}$ is the set of time points that depends on the clonotype of interest and $\sigma_G^{data}(t)$ represents the standard deviation of the experimental data at time $t$. 

When refitting the data using the hypoexponential model, we estimated the mean and variance of the duration of the first and subsequent generations, $(\tau_{G=0},\sigma^2_{G=0})$ and $(\tau_{G>0},\sigma^2_{G>0}),$ which is equivalent to estimating the rates $(\lambda_0, \lambda)$ and number of stages ($N_0, N)$ in the Erlang formulation.

\subsubsection{Chemotherapy induced neutropenia}

The \citet{Friberg2002} semi-mechanistic model of chemotherapy induced neutropenia includes four kinetic parameters to describe the dynamics of neutrophil precursors and circulating neutrophils, $k_p, k_{circ}, N_0,$ and $ \gamma,$ with two further parameters to characterise, $E_{max}$ and $EC_{50}$, to capture the effect of anti-cancer chemotherapy on proliferating neutrophil precursors. Here, we use the population PK parameters for Zalypsis found by \citet{Perez-Ruixo2012}. 

Finally, we must also estimate two distribution-specific parameters. In the case of Erlang distributed  maturation durations, these parameters are given by the mean maturation delay, $\tau_M$, and the number of maturation compartments, $M$. In this case, we estimate $\theta_{Neutropenia} = [k_p, k_{circ}, N_0,\gamma, E_{max},EC_{50}, \tau_{M}, M]$. Conversely, in the case of hypoexponential distributed maturation durations, we estimate the mean maturation delay $\tau_M$ and variance $\sigma^2_M$, so $\theta_{Neutropenia} = [k_p, k_{circ}, N_0,\gamma, E_{max},EC_{50}, \tau_{M}, \sigma_{M}^2]$. Thus, when using either the hypoexponential or Erlang distributions, there are eight total parameters to be fit to the simulated neutrophil data. 

Then, we fit the model parameters for both the Erlang and hypoexponential distributed maturation times by minimizing
\begin{align} \label{Eq:FribergMaximumLikelihood}
    SSR(\theta_{Neutropenia} ) =  \sum_{i=1}^{21} \left[  \log_{10} ( N(t_{i},\theta_{Neutropenia}) ) -\log_{10}(N_{i}^{data})  \right]^2  , 
\end{align}
where $N(t_{i},\theta_{Neutropenia})$ is the simulated neutrophil concentration at time $t_i.$ We consider the simulated neutrophil data on $\log_{10}$ scale and make the common assumption of Gaussian error on this scale so that minimizing Eq.~\eqref{Eq:FribergMaximumLikelihood} corresponds to maximizing the log-likelihood. For both the Erlang and hypoexponential distributions, we minimized Eq.~\eqref{Eq:FribergMaximumLikelihood} using the genetic algorithm $\texttt{ga}$ implemented in \citet{Matlab2017}. When considering the hypoexponential distributed maturation duration, we estimated all eight real-valued model parameters simultaneously. Conversely, for the Erlang distributed maturation duration, we consider the mixed-optimization problem for the integer-valued shape parameter $M$ and the remaining seven real-valued model parameters. 

\section{Results}

\subsection{Eclipse phase duration in SHIV}

\begin{figure}[htbp] 
\centering  
\includegraphics[scale=0.87, trim= 10 10 10 10,clip]{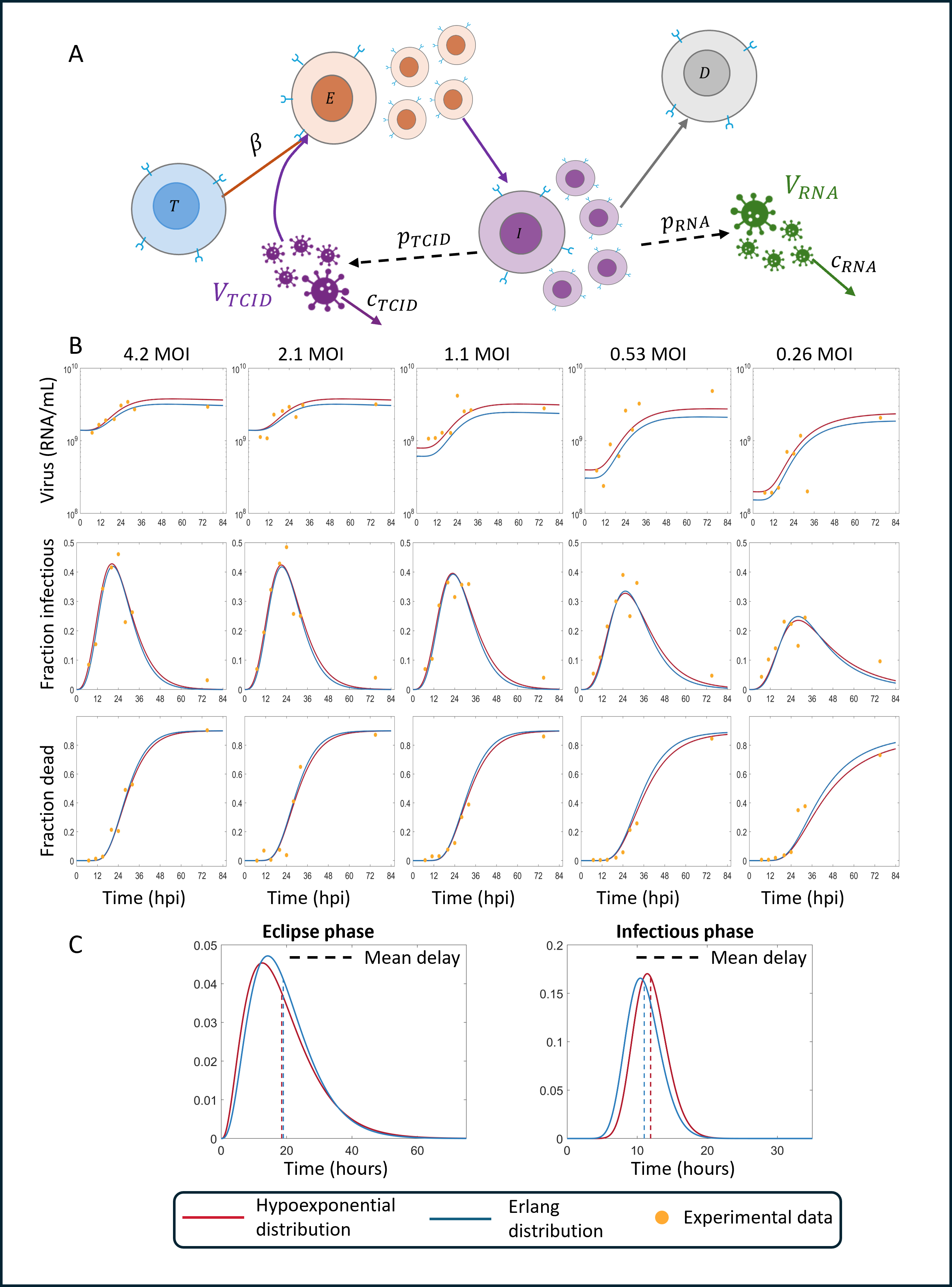}    
\caption{\textbf{A)} Schematic representation of the \citet{Beauchemin2017} model.  \textbf{B)} Model fits to the \citet{Beauchemin2017} data. The experimental data across five MOIs is plotted as filled circles and the model predictions are continuous curves. The red curve is the model fit using the hypoexponential distribution and the blue curve is the model fit using the Erlang distribution. \textbf{C)} Comparison of the distribution PDFs for the duration of the eclipse and infectious phase of infection. The dashed vertical lines give the mean of each distribution}
 
\label{Fig:BeaucheminFigure}
\end{figure}

As mentioned, the digitized data that we fit  does not substantially differ from the data fit by \citet{Beauchemin2017}. Consequently, we can compare our fitting results with their best-fit results. We fit the hypoexponentially distributed eclipse phase model Eq.~\eqref{Eq:BeaucheminHypoexponential} to this data from 10 distinct initial guesses and show the model fits to the digitized data in Fig.~\ref{Fig:BeaucheminFigure}B. There, we compare model fits to the digitized data using the best-fit parameters from \citet{Beauchemin2017} against the model fits using Eq.~\eqref{Eq:BeaucheminHypoexponential}. 

In general, both models fit the data well across the five different MOIs. However, as measured by the log-likelihood Eq.~\eqref{Eq:BeaucheminMaximumLikelihood} and AIC, the hypoexponential model fits the data significantly better than the Erlang approximation used by \citet{Beauchemin2017}, with AIC values of 114.6 and 128.6, respectively. This improved fit is primarily driven by the improved fit to the virus RNA measurements, shown in the first row of Fig.~\ref{Fig:BeaucheminFigure}B. As these two models have the same number of free parameters, the improved AIC for the hypoexponential model is directly due to the improved fit to the experimental data. Indeed, the value of Eq.~\eqref{Eq:BeaucheminMaximumLikelihood} is roughly 15 points lower than the best fit found by \citet{Beauchemin2017}, which indicates strong support for the hypoexponential fit \citep{Burnham2004}.  Furthermore, the hypoexponential model can fit all model parameters simultaneously and avoid the mixed-optimization problem that \citet{Beauchemin2017} noted ``cause(d) the fitter to misbehave.'' 

In Fig.~\ref{Fig:BeaucheminFigure}C, we compare the PDFs of the inferred durations of the eclipse and infectious phases from the Erlang and hypoexponential fitting. \citet{Beauchemin2017} estimated the mean duration of the eclipse and infectious phases as 19 and 11 hours. Here, when using the hypoexponential distribution, we estimate a mean duration of 18.5 and 11.9 hours for these two processes, respectively. While the inferred means were similar, the standard deviation of the inferred distribution of the duration of the eclipse phase differed between the Erlang and hypoexponential fitting, 9.5 hours and 10.3 hours, respectively. This difference is apparent in the PDFs in Fig.~\ref{Fig:BeaucheminFigure}C, with increased heterogeneity in the duration of the eclipse phase obtained using the hypoexponential model. Conversely, both models estimate that the duration of the infectious phase is more homogeneous, with smaller standard deviations of 2.5 and 2.4 hours, respectively. However, the hypoexponential model predicts that cells spend slightly longer in the infectious phase of infection, which allows the model to capture the slower accumulation of dead cells while capturing the increased viral production in experiments with high MOI. As mentioned, the improvement in fit for the hypoexponential model is primarily driven by these improved fits to virus RNA. Ultimately, as shown in Table~\ref{Table:BeaucheminParameterTable}, the estimated parameters were similar between the hypoexponential and Erlang distributions.

\begin{table}[htbp!] 
\begin{tabular}{l|l|l} \sphline
\textbf{Parameter}    (units)      & \textbf{Erlang}   & \textbf{Hypoexponential}        \\ 
$\beta$ (mL/TCID$_{50}$/hr) 	& $10^{-5.49}$ & $10^{-5.68}$    \\ \hline 
$p_{RNA}$ (vRNA/cell/hr) 	& $10^{2.31}$ & $10^{2.39}$    \\ \hline  
Base MOI(TCID$_{50}$/Cell)	& 0.41 & 0.53     \\ \hline 
$\tau_{E}$ (hours)	& 19 & 18.5   \\ \hline 
$\sigma^2_{E}$ (hours$^2$)	& 90 & 107    \\ \hline 
$\tau_{I}$ (hours)	& 11 & 11.9     \\ \hline 
$\sigma^2_{I}$ (hours$^2$)	& 6.1 & 5.7   \\ \hline 
\end{tabular}
\caption{ Best fit parameter estimates from fitting Eq.~\eqref{Eq:BeaucheminDDE} assuming either Erlang distributions or hypoexponential distributions for the duration of the eclipse and infectious phases. The best-fit parameters in the eclipse phase are taken from \citet{Beauchemin2017} and transformed to an hourly timescale.   }
\label{Table:BeaucheminParameterTable}
\end{table}

\subsubsection*{Implementation of the hypoexponential approach within a Bayesian inference framework}

\begin{figure}[h!]
\centering
\includegraphics[width=1\textwidth]{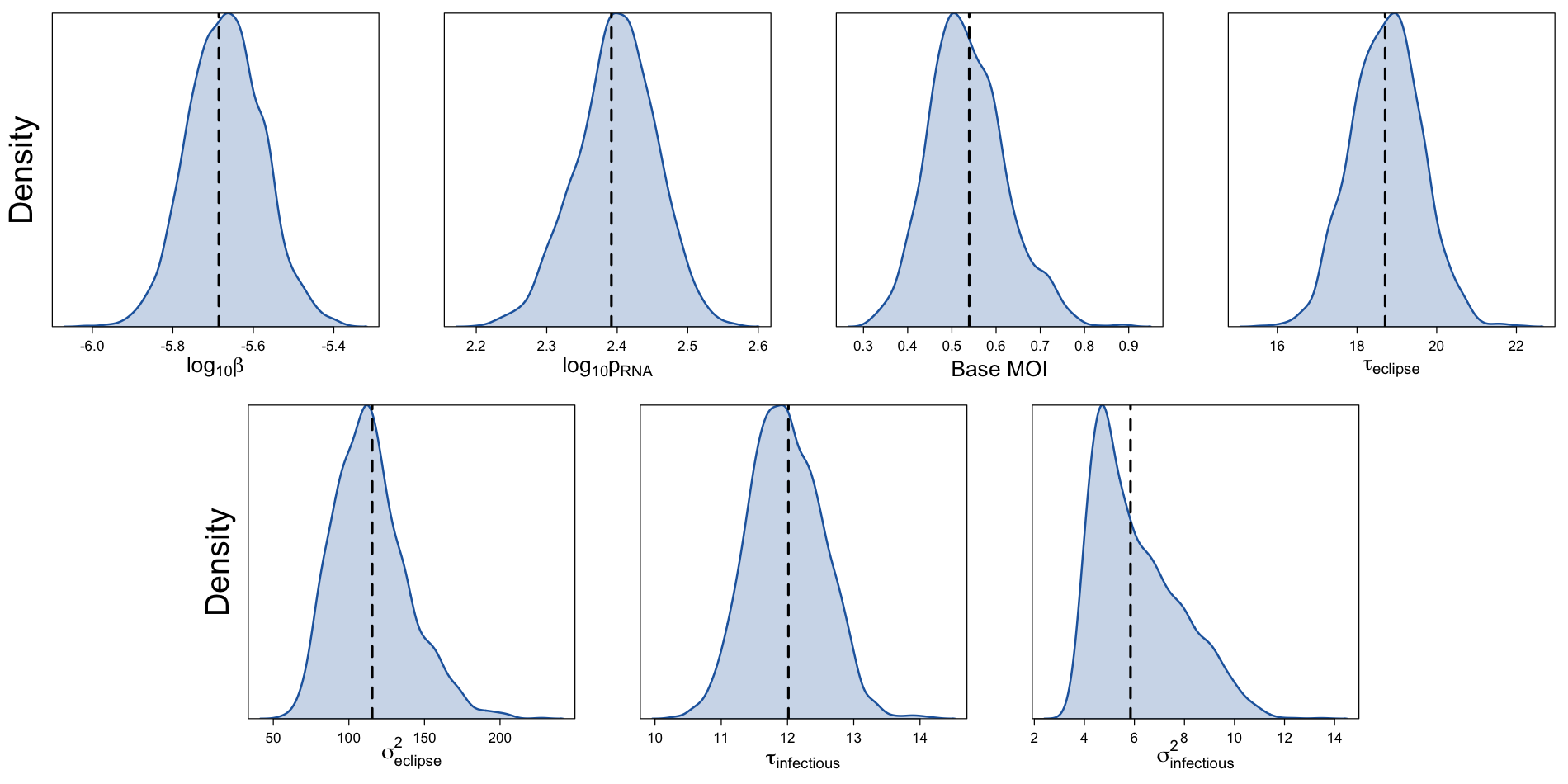}
\caption{
Posterior distributions of the inferred kinetic parameters in Eq.~\eqref{Eq:BeaucheminHypoexponential} obtained by implementing the hypoexponential model in Stan. The point estimates in Table~\ref{Table:BeaucheminParameterTable}  are overlaid on each distribution as dashed lines.}
\label{Fig:stan_posteriors}
\end{figure}

We refit the SHIV model (Eq.~\eqref{Eq:BeaucheminHypoexponential}) in the probabilistic programming framework Stan~\citep{Carpenter2017}, which builds on Hamiltonian Monte Carlo (HMC). We thus obtain a posterior distribution of the biologically meaningful parameters $\tau_{E}, \sigma^2_{E},\tau_{I}$, and $\sigma^2_{I}$. We sample  $j_E$ and $j_I$ uniformly over the biologically plausible ranges $j_E \in [1,6]$, and $j_I \in [14,35]$, as excessively wide ranges for these parameters can slow or prevent convergence. These ranges for $j_I$ and $j_E$ also fix the possible number of ODEs required to implement the hypoexponential model in Stan and we provide the corresponding Stan code. We used weakly informative normal priors for the remaining parameters centered on the point estimates in Table \ref{Table:BeaucheminParameterTable}. We show the resulting posteriors for the kinetic parameters in Fig.~\ref{Fig:stan_posteriors} and the resulting model dynamics for different MOIs as a Bayesian equivalent of Fig.~\ref{Fig:BeaucheminFigure}B in Fig.~\ref{Fig:stan_dynamics}.

\subsection{Division times in lymphocyte proliferation}

We next fit the generation dynamics of both OT-I and F5 cells reported by \citet{Belluccini2022}. As \citet{Belluccini2022} fit their model using SMC to obtain posterior parameter distributions, we cannot directly compare our modelling fits against their results. Consequently, we re-fit the OT-I and F5 data using the Erlang models for the 36 pairs of $N_0 = 2,3,4,5,6,7$ and $N = 2,3,4,5,6,7,$ which results in 36 optimization problems to infer the remaining model parameters $(C_0, \lambda_0, \lambda, \alpha)$. For both the OT-I and F5 data, these 36 optimization problems cumulatively took over $3\times 10^5$ seconds on a personal laptop. While individual optimization problems for a pair $(N_0,N)$ of the Erlang model estimates fewer parameters than the hypoexponential model, the need to test multiple pairs of $N$ and $N_0$ creates computational inefficiencies. Conversely, when fitting the hypoexponential model to this data, we have a single optimization problem to infer the six parameters $(C_0, \lambda_0, \lambda, \alpha, \sigma^2_0, \sigma^2 )$. For the OT-I and F5 data, these optimization problems took $1 \times 10^4$ and $3 \times 10^4$ seconds on the same personal computer, respectively. Consequently, using the hypoexponential model, rather than the Erlang model with the resulting mixed-optimization problem, results in a 10-fold computational time savings.  

\begin{figure}
    \centering
    \includegraphics[scale=0.87, trim= 10 5 10 10,clip]{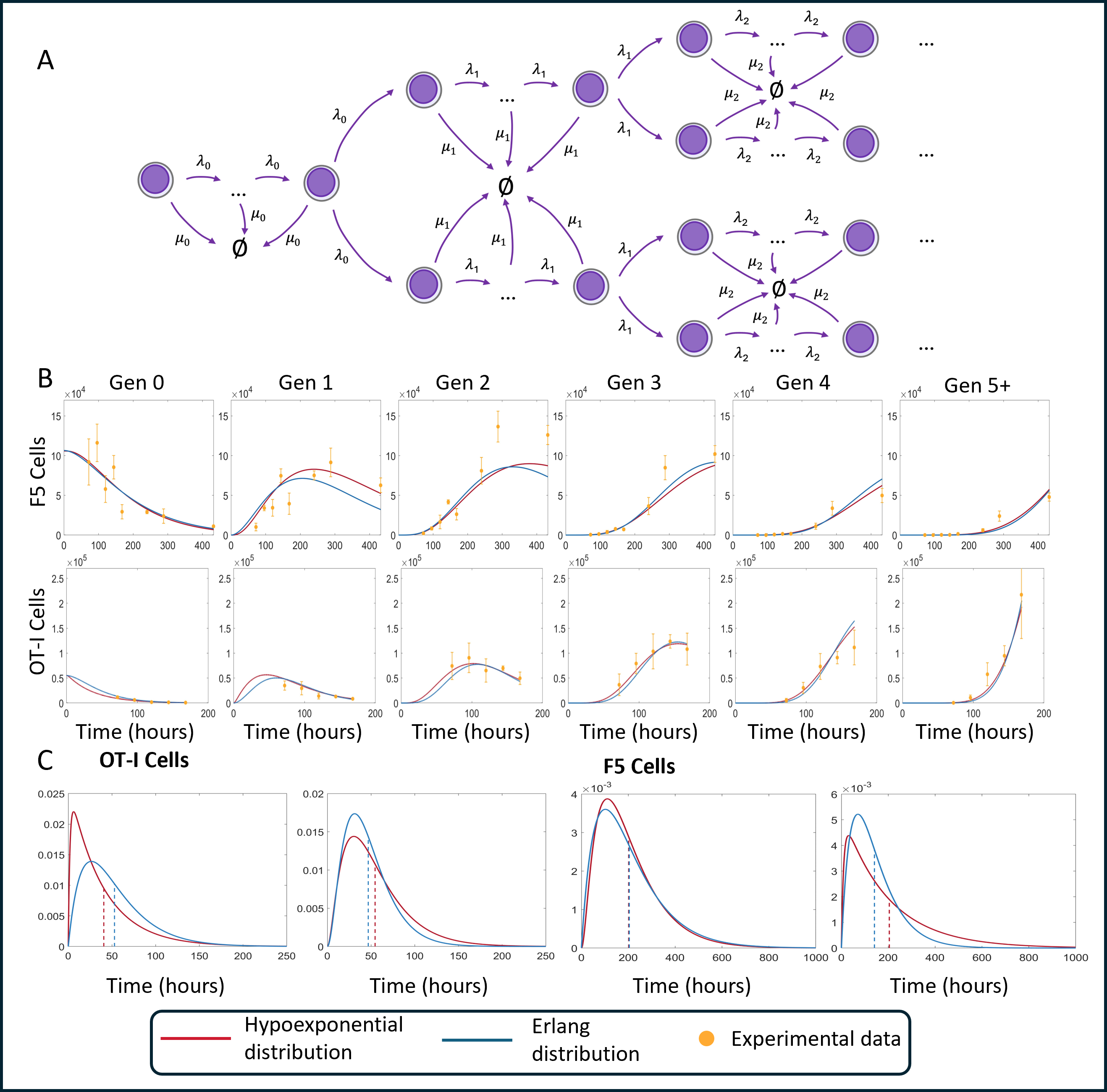} 
    \caption{\textbf{A)} Schematic representation of the \citet{Belluccini2022} model.  \textbf{B)} Model fits to the \citet{Belluccini2022} data. The experimental data are plotted as filled circles, with the error bars representing the standard deviation of the experimental data. The continuous red and blue curves show the model fits using the hypoexponential and Erlang distribution, respectively. \textbf{C)} Comparison of the distribution PDFs for the duration of the generation time for cells in the first and later generations, respectively. The dashed vertical lines give the mean of each distribution.}
    \label{Fig:BellucciniFig}
\end{figure}

Both the hypoexponential and the Erlang models fit the data well, with the model fits for each cell-line given in Fig.~\ref{Fig:BellucciniFig}. The AICs for the hypoexponential and Erlang models are 22.9 and 25.3 for OT-I data and 37.9 and 40.5  for the F5 data, respectively. For both cell lines, the hypoexponential model fit the data better, and the difference in AIC values indicates moderate support for the hypoexponential model over the Erlang model \citep{Burnham2004}. As these two models have the same number of free parameters to be inferred from the data, the improved AIC for the hypoexponential model is directly due to the improved fit to the experimental data. Consequently, the hypoexponential model is both more computationally efficient and parsimonious with the experimental data than the Erlang model.

 Further, our grid search of possible $(N_0,N)$ pairs for the Erlang model corresponds to the profile likelihood with respect to these integer-valued parameters \citep{Raue2009}. For the OT-I data, this discrete profile likelihood indicates that 33 of the $(N_0,N)$ pairs fall within the 95$\%$ confidence interval for these parameters. This demonstrates that the profile likelihood surface of these two parameters is relatively flat and suggests practical non-identifiability of these discrete parameters. Indeed, the only pairs that do not fall within the 95$\%$ confidence interval  have $N_0 = 7$ and $N \geq 5$. Similarly, 23 of the 36 $(N_0,N)$ pairs fall within the 95$\%$ confidence interval for the F5 data, which again suggests practical non-identifiability of these discrete parameters. 

We report the estimated parameters in Table~\ref{Table:BellucciniParameterTable} and note the significant difference between the hypoexponential and Erlang distribution times.  Unlike the \citet{Beauchemin2017} example, where the inferred distributions were similar, the hypoexponential and Erlang models have significantly different distributions for the inter-generation time. We compare the PDFs of the inferred generation times for the 0th and later generations in Fig.~\ref{Fig:BellucciniFig}C. Here, the increased flexibility of using the hypoexponential model instead of the Erlang model is particularly apparent in the inferred distributions for the duration of the 0th generation of OT-I cells and in the distribution of the 0th and later generations for F5 cells. In both cases, the mean estimates for the duration of these generations are similar, but the resulting heterogeneity in the inferred distributions differ significantly. The decreased heterogeneity in the duration of the 0th generation of OT-I cells is apparent in the rapid accumulation of OT-I cells in the 1st generation. The maximal concentration of 1st generation OT-I cells is predicted to occur much earlier in the hypoexponential model than the Erlang model. However, this difference becomes less pronounced for later generations, as the Erlang model predicts a slightly shorter duration of these later generations. As the other parameters in these models, $C_0$ and $\alpha$, are extremely similar between the hypoexponential and Erlang models, the improvement in fit and the differences in the predicted dynamics are strictly due to the increased flexibility in capturing the duration of each generation. Our parameter estimation predicts that the 0th generation has a shorter mean duration than the later generations for both hypoexponential models. However, for F5 cells, the median generation duration is shorter for the later distributions, which is consistent with other biological evidence \citep{Belluccini2022}, due to the high variance in this distribution.

\begin{table}[htbp!] 
\begin{tabular}{l|l|l|l|l} \sphline
 \textbf{Parameter} (units)   & \multicolumn{2}{l|}{\textbf{OT-I cells}}       & \multicolumn{2}{l}{\textbf{F5 cells}}     \\ 
     &  Erlang model &   Hypoexponential model &  Erlang model &   Hypoexponential model \\ \hline 
$C_0$ (cells) 					&  $10^{4.7}$  & $10^{4.7}$ 				&  $10^{5.0}$  & $10^{5.0}$    \\ \hline 
$\alpha$ (1/hr) 				&  $10^{-3.1}$ & $10^{-3.1}$ 				&  $10^{-2.9} $ & $10^{-2.9}$    \\ \hline    
$\tau_{G=0}$ (hours)			&  52.9 & 40.7  			  		  &  204 & 202    \\ \hline 
$\sigma^2_{G = 0}$ (hours$^2$)	&  1.4$\times 10^3$   & 1.5$\times 10^3$  		  &   2.1$\times 10^4$ & 1.9$\times 10^4$   \\ \hline 
$\tau_{G > 0}$ (hours) 			&  46.8 & 54.6  					  &  141 & 204    \\ \hline 
$\sigma^2_{G > 0}$ (hours$^2$)	&  729  & 1.3$\times 10^3$ &  9.9$\times 10^3$ & 3.8$\times 10^4$ \\ \hline 
\end{tabular}
\caption{ Best fit parameter estimates from fitting Eq.~\eqref{Eq:LymphocyteGenerationsDDE} assuming either Erlang distributions or hypoexponential distributions for OT-I and F5 generation data. We calculate the mean generation duration using the inferred values of $\lambda$ and $\lambda_0$ for each distribution.   }
\label{Table:BellucciniParameterTable}
\end{table}
 
\subsection{Chemotherapy induced neutropenia}
Finally, we fit the Friberg model to the synthetic neutropenia data using a genetic algorithm using both the Erlang and hypoexponential distributions. This example illustrates the benefits of the hypoexponential approach in a canonical transit-compartment PK/PD model, rather than as an additional validation against experimental data. Both the Erlang and hypoexponential models fit the data well, with a negligible difference in the SSR values between the Erlang and hypoexponential models ($0.14$ and $0.13$, respectively). However, the hypoexponential model was over 4-fold more computationally efficient, with runtimes of $6.7 \times 10^4$ and $1.6\times 10^4$ seconds, for the Erlang and hypoexponential model, respectively. While the difference between model fits is negligible, the hypoexponential model captures the neutrophil nadir better than the Erlang model, although both models over-estimate the nadir concentrations. We report the estimated parameter values in Table~\ref{Table:FribergParameterTable}.
 
As in the SHIV example, the inferred distribution of the maturation durations is similar between the Erlang and hypoexponential models. However, the hypoexponential distribution has a slightly larger variance that indicates increased heterogeneity in the duration of the maturation lag; the inferred mean and variance do not satisfy the implicit relationship of an Erlang distribution and thus illustrate how using a hypoexponential distribution can capture more general biological dynamics. Indeed, a gamma distribution with the same mean and variance as the hypoexponential distribution would have shape parameter $ j =3.9 $. Conversely, the best fit value for the Erlang model was $j = 5$, although the $j=4$ and $j =6$ cases were comparable. This suggests that the implicit relationship between the mean and variance of an Erlang distribution results in potential identifiability issues in estimating the heterogeneity in the duration of maturation, as models with different numbers of compartments may equally well approximate the underlying delayed process. 

Overall, we found that the goodness-of-fit of the Erlang model decreases as the number of compartments $M$ increases, with the corresponding decrease in the heterogeneity in the duration of neutrophil maturation. However, small values of $M$, with corresponding higher variance, enforce a decrease in the mean maturation time that led to a further underestimation of the nadir concentration. Conversely, the hypoexponential model decouples the mean and variance and allows for more flexibility in inferring the underlying distribution. 
    
\begin{table}[htbp!] 
\begin{tabular}{l|l|l} \sphline
\textbf{Parameter}    (units)      & \textbf{Erlang}   & \textbf{Hypoexponential}        \\ \hline
$k_p$ (1/day) 	&  0.33 & 0.27    \\ \hline 
$E_{max}$ (unitless) & 0.95  & 0.98    \\ \hline  
$EC_{50}$ (ng/mL)	& 2.07 & 1.02      \\ \hline 
$\gamma $ (unitless)	& 0.53 & 0.71      \\ \hline
$N_0$ ($10^9$ cells/L)  & 5.37 & 5.27      \\ \hline 
$k_{circ} $ (1/day) 	& 1.54 & 1.75      \\ \hline  
$\tau_{M}$ (days)	& 3.25 & 3.32   \\ \hline 
$\sigma^2_{M}$ (days$^2$) & 2.11 & 3.11    \\ \hline  
\end{tabular}
\caption{ Best fit parameter estimates from fitting Eq.~\eqref{Eq:FribergDDE} assuming either Erlang distributions or hypoexponential distributions to the synthetic neutropenia data.    }
\label{Table:FribergParameterTable}
\end{table}

\section{Discussion}

Transit compartment models are extensively used to capture the effects of physiological delays across the pharmaceutical sciences \citep{Hof2021,Koch2014,deSouza2017,Bauer2025}. These transit compartments are equivalent to Erlang distributions through the well-known linear chain technique that allows modellers to capture the physiologically relevant delays while retaining an equivalent ODE representation. However, the standard transit compartment formulation implicitly imposes an artificial relationship between the mean and variance of the delayed process \citep{Cassidy2022}. Here, we generalize this transit compartment modelling framework to remove this implicit assumption and retain the computational efficiency and simplicity of an equivalent ODE representation using hypoexponential distributed DDEs.

These hypoexponential distributed DDEs are a natural extension of the common transit compartment models; they differ only in allowing the rates between compartments to vary. Consequently, they are straightforward to implement for modellers who are familiar with transit compartment models. Furthermore, the rates of the hypoexponential distribution are explicitly determined by the mean and variance of the delayed process. The resulting characterization of the delayed process may facilitate understanding of the underlying biological mechanism. Furthermore, Erlang distributed DDEs are nested models of the more general hypoexponential distribution. Consequently, the hypoexponential distributed models will fit experimental data at least as well as the common Erlang models. Indeed, we have shown through three examples that using a hypoexponential model rather than the Erlang model leads to improved fits to data from a variety of settings. In the \citet{Beauchemin2017} and \citet{Belluccini2022} examples, there was statistical support for the hypoexponential model over the Erlang model. 

These examples also illustrate the increased computational efficiency of the hypoexponential model compared to the Erlang models. As Erlang models must have integer-valued shape parameters, estimating the number of compartments results in a mixed-optimization problem that is typically computationally demanding. Moreover, this implicit relationship between the mean and variance of the delayed process can artificially bias parameter inference. Conversely, as the rates of the hypoexponential distribution are explicitly determined by the mean and variance, modellers can directly estimate these real-valued quantities without enforcing an artificial relationship. Moreover, directly estimating these quantities when using hypoexponential distributed models avoids the mixed-optimization problem that is inherent to standard transit compartment models. Accordingly, there was a significant decrease in computational cost when using the hypoexponential distribution. Ultimately, as the models considered in this work have been used extensively elsewhere \citep{Quirouette2023,Bernhauerova2021,Soto2020}, our results illustrate the potential to improve model fits and predictions and minimize computational cost by using a simple and straightforward extension of the common Erlang distribution that removes artificial assumptions on the delay while preserving the ability to utilize efficient numerical methods for ODEs. In particular, due to the historical formulation of the model, most applications consider three compartments when using the Friberg model, and do not investigate the sensitivity of model predictions to this modelling choice due to the resulting computational cost \citep{Hu2018,Krzyzanski2018}. Here, we have shown how to relax this assumption while retaining the computational efficiency of existing software packages. 

There has been increased interest in adapting existing software packages to allow modellers to use more general distributed DDEs \citep{Hu2018,Bauer2025,Guglielmi2024,Cassidy2024}. For example, \citet{Krzyzanski2018} used the convolution operator in Phoenix to compare a gamma-distributed DDE formulation of the Friberg model against the common Erlang formulation. As in our third example, \citet{Krzyzanski2018} observed improved model fits when using the more general distributed DDE, but this flexibility imposed a significant computational cost as the distributed DDE model was roughly 3000-fold slower than the ODE implementation of the Erlang model \citep{Krzyzanski2018}. Here, we have achieved both improved model fits and increased computational efficiency by using the hypoexponential, rather than gamma, distributed DDEs. 

While hypoexponential distributed DDEs can be implemented as a system of ODEs with an identical structure to the familiar transit compartment models, we note that a package implementation of the hypoexponential distributed DDE is not currently available in most software packages. However, we expect the implementation of this approach to be straightforward in these packages and we implemented this approach in Stan for the SHIV example. Further, we have not investigated the identifiability of the model parameters in these hypoexponential distributed DDEs. However, in the case of hypoexponential distributed DDEs, these parameters are all real, rather than integer, valued which facilitates the use of existing software packages to investigate parameter identifiability or perform Bayesian inference, as we have shown \citep{Bellu2007,Cassidy2023a,Raue2009,Porthiyas2024}. Despite these limitations, we have presented a simple to implement generalization of Erlang transit compartment models that removes the need for integer parameter estimation, preserves the computational efficiency of equivalent ODE formulations, and improves inference of delayed processes across QSP applications.


\section*{Code availability}

The code and experimental data used in this study are available at \\ 
\texttt{https://github.com/ttcassid/EfficientTransitCompartmentModels/}. 



\begin{thebibliography}{}

\bibitem[Anderson and Burnham, 2004]{anderson2004model}
Anderson, D. and Burnham, K. (2004).
\newblock Model selection and multi-model inference.
\newblock {\em Springer-Verlag}, 63(2020):10.

\bibitem[Bauer and Krzyzanski, 2025]{Bauer2025}
Bauer, R.~J. and Krzyzanski, W. (2025).
\newblock {Tutorial for Modeling Delays in Biological Systems in the NONMEM
  Software}.
\newblock {\em CPT Pharmacometrics Syst. Pharmacol.}, 14(7):1133--1155.

\bibitem[Beauchemin et~al., 2008]{Beauchemin2008}
Beauchemin, C.~A., McSharry, J.~J., Drusano, G.~L., Nguyen, J.~T., Went, G.~T.,
  Ribeiro, R.~M., and Perelson, A.~S. (2008).
\newblock {Modeling amantadine treatment of influenza A virus in vitro}.
\newblock {\em J. Theor. Biol.}, 254(2):439--451.

\bibitem[Beauchemin et~al., 2017]{Beauchemin2017}
Beauchemin, C.~A., Miura, T., and Iwami, S. (2017).
\newblock {Duration of SHIV production by infected cells is not exponentially
  distributed: Implications for estimates of infection parameters and antiviral
  efficacy}.
\newblock {\em Sci. Rep.}, 7(February).

\bibitem[Bellu et~al., 2007]{Bellu2007}
Bellu, G., Saccomani, M.~P., Audoly, S., and D'Angi{\`{o}}, L. (2007).
\newblock {DAISY: A new software tool to test global identifiability of
  biological and physiological systems}.
\newblock {\em Comput. Methods Programs Biomed.}, 88(1):52--61.

\bibitem[Belluccini et~al., 2022]{Belluccini2022}
Belluccini, G., L{\'{o}}pez-Garc{\'{i}}a, M., Lythe, G., and
  Molina-Par{\'{i}}s, C. (2022).
\newblock {Counting generations in birth and death processes with competing
  Erlang and exponential waiting times}.
\newblock {\em Sci. Rep.}, 12(1):11289.

\bibitem[Bernhauerov{\'{a}} et~al., 2021]{Bernhauerova2021}
Bernhauerov{\'{a}}, V., Lisowski, B., Rezelj, V.~V., and Vignuzzi, M. (2021).
\newblock {Mathematical modelling of SARS-CoV-2 infection of human and animal
  host cells reveals differences in the infection rates and delays in viral
  particle production by infected cells}.
\newblock {\em J. Theor. Biol.}, 531.

\bibitem[Burnham and Anderson, 2004]{Burnham2004}
Burnham, K.~P. and Anderson, D.~R. (2004).
\newblock {Multimodel Inference}.
\newblock {\em Sociol. Methods Res.}, 33(2):261--304.

\bibitem[{C{\^{a}}mara De Souza} et~al., 2018]{deSouza2017}
{C{\^{a}}mara De Souza}, D., Craig, M., Cassidy, T., Li, J., Nekka, F.,
  B{\'{e}}lair, J., and Humphries, A.~R. (2018).
\newblock {Transit and lifespan in neutrophil production: implications for drug
  intervention}.
\newblock {\em J. Pharmacokinet. Pharmacodyn.}, 45(1):59--77.

\bibitem[Carpenter et~al., 2017]{Carpenter2017}
Carpenter, B., Gelman, A., Hoffman, M.~D., Lee, D., Goodrich, B., Betancourt,
  M., Brubaker, M., Guo, J., Li, P., and Riddell, A. (2017).
\newblock {Stan : A Probabilistic Programming Language}.
\newblock {\em J. Stat. Softw.}, 76(1).

\bibitem[Cassidy, 2021]{Cassidy2020a}
Cassidy, T. (2021).
\newblock {Distributed Delay Differential Equation Representations of Cyclic
  Differential Equations}.
\newblock {\em SIAM J. Appl. Math.}, 81(4):1742--1766.

\bibitem[Cassidy, 2023]{Cassidy2023a}
Cassidy, T. (2023).
\newblock {A continuation technique for maximum likelihood estimators in
  biological models}.
\newblock {\em Bull. Math. Biol.}, 85(10):90.

\bibitem[Cassidy, 2025]{Cassidy2024}
Cassidy, T. (2025).
\newblock {Using Multidelay Discrete Delay Differential Equations to Accurately
  Simulate Models With Distributed Delays}.
\newblock {\em Stud. Appl. Math.}, 154(6):1--18.

\bibitem[Cassidy et~al., 2019]{Cassidy2018a}
Cassidy, T., Craig, M., and Humphries, A.~R. (2019).
\newblock {Equivalences between age structured models and state dependent
  distributed delay differential equations}.
\newblock {\em Math. Biosci. Eng.}, 16(5):5419--5450.

\bibitem[Cassidy et~al., 2022]{Cassidy2022}
Cassidy, T., Gillich, P., Humphries, A.~R., and van Dorp, C.~H. (2022).
\newblock {Numerical methods and hypoexponential approximations for gamma
  distributed delay differential equations}.
\newblock {\em IMA J. Appl. Math.}, 87(6):1043--1089.

\bibitem[Cassidy and Humphries, 2019]{Cassidy2018}
Cassidy, T. and Humphries, A.~R. (2019).
\newblock {A mathematical model of viral oncology as an immuno-oncology
  instigator}.
\newblock {\em Math. Med. Biol. A J. IMA}, 37(1):117--151.

\bibitem[Champredon et~al., 2018]{Champredon2018}
Champredon, D., Dushoff, J., and Earn, D. J.~D. (2018).
\newblock {Equivalence of the Erlang-Distributed SEIR Epidemic Model and the
  Renewal Equation}.
\newblock {\em SIAM J. Appl. Math.}, 78(6):3258--3278.

\bibitem[Craig, 2017]{Craig2017}
Craig, M. (2017).
\newblock {Towards Quantitative Systems Pharmacology Models of
  Chemotherapy-Induced Neutropenia}.
\newblock {\em CPT Pharmacometrics Syst. Pharmacol.}, 6(5):293--304.

\bibitem[Craig et~al., 2016]{Craig2016}
Craig, M., Humphries, A.~R., and Mackey, M.~C. (2016).
\newblock {A mathematical model of granulopoiesis incorporating the negative
  feedback dynamics and kinetics of G-CSF/neutrophil binding and
  internalization}.
\newblock {\em Bull. Math. Biol.}, 78(12):2304--2357.

\bibitem[Diekmann and Gyllenberg, 2012]{Diekmann2012}
Diekmann, O. and Gyllenberg, M. (2012).
\newblock {Equations with infinite delay: Blending the abstract and the
  concrete}.
\newblock {\em J. Differ. Equ.}, 252(2):819--851.

\bibitem[Diekmann et~al., 2018]{Diekmann2017}
Diekmann, O., Gyllenberg, M., and Metz, J. A.~J. (2018).
\newblock {Finite dimensional state representation of linear and nonlinear
  delay systems}.
\newblock {\em J. Dyn. Differ. Equations}, 30(4):1439--1467.

\bibitem[Diekmann et~al., 2020a]{Diekmann2020}
Diekmann, O., Gyllenberg, M., and Metz, J. A.~J. (2020a).
\newblock {Finite dimensional state representation of physiologically
  structured populations}.
\newblock {\em J. Math. Biol.}, 80(1-2):205--273.

\bibitem[Diekmann et~al., 2020b]{Diekmann2020a}
Diekmann, O., Gyllenberg, M., and Metz, J. A.~J. (2020b).
\newblock {On models of physiologically structured populations and their
  reduction to ordinary differential equations}.
\newblock {\em J. Math. Biol.}, 80(1-2):189--204.

\bibitem[Diekmann et~al., 2012]{Diekmann2012a}
Diekmann, O., Heesterbeek, H., and Britton, T. (2012).
\newblock {\em {Mathematical Tools for Understanding Infectious Disease
  Dynamics}}.
\newblock Princeton University Press.

\bibitem[Diekmann et~al., 2020c]{Diekmann2020c}
Diekmann, O., Scarabel, F., and Vermiglio, R. (2020c).
\newblock {Pseudospectral discretization of delay differential equations in
  sun-star formulation: Results and conjectures}.
\newblock {\em Discret. Contin. Dyn. Syst. - S}, 13(9):2575--2602.

\bibitem[Di{\'{e}}ras et~al., 2019]{Dieras2019}
Di{\'{e}}ras, V., Harbeck, N., Joy, A.~A., Gelmon, K., Ettl, J., Verma, S., Lu,
  D.~R., Gauthier, E., Schnell, P., Mori, A., Rugo, H.~S., and Finn, R.~S.
  (2019).
\newblock {Palbociclib with Letrozole in Postmenopausal Women with ER+/HER2−
  Advanced Breast Cancer: Hematologic Safety Analysis of the Randomized
  PALOMA-2 Trial}.
\newblock {\em Oncologist}, 24(12):1514--1525.

\bibitem[Friberg et~al., 2002]{Friberg2002}
Friberg, L.~E., Henningsson, A., Maas, H., Nguyen, L., and Karlsson, M.~O.
  (2002).
\newblock {Model of chemotherapy-induced myelosuppression with parameter
  consistency across drugs}.
\newblock {\em J. Clin. Oncol.}, 20(24):4713--4721.

\bibitem[Ganusov et~al., 2007]{ganusov20072}
Ganusov, V.~V., Milutinovic, D., and De~Boer, R.~J. (2007).
\newblock Il-2 regulates expansion of CD4+ T cell populations by affecting cell
  death: insights from modeling CFSE data.
\newblock {\em J. Immunol}, 179(2):950--957.

\bibitem[Guglielmi and Hairer, 2025]{Guglielmi2024}
Guglielmi, N. and Hairer, E. (2025).
\newblock {Applying Stiff Integrators for Ordinary Differential Equations and
  Delay Differential Equations to Problems with Distributed Delays}.
\newblock {\em SIAM J. Sci. Comput.}, 47(1):A102--A123.

\bibitem[Hale and {Verduyn Lunel}, 1993]{Hale1993}
Hale, J.~K. and {Verduyn Lunel}, S.~M. (1993).
\newblock {Introduction to Functional Differential Equations}.
\newblock {\em Appl. Math. Sci.}, 99.

\bibitem[Hawkins et~al., 2007]{hawkins2007model}
Hawkins, E.~D., Turner, M.~L., Dowling, M.~R., Van~Gend, C., and Hodgkin, P.~D.
  (2007).
\newblock A model of immune regulation as a consequence of randomized
  lymphocyte division and death times.
\newblock {\em Proc. Natl. Acad. Sci. U.S.A.},
  104(12):5032--5037.

\bibitem[Hino et~al., 1991]{Hino1991}
Hino, Y., Murakami, S., and Naito, T. (1991).
\newblock {Functional Differential Equations with Infinite Delay}.
\newblock 1473.

\bibitem[Hof and Bridge, 2021]{Hof2021}
Hof, F. and Bridge, L.~J. (2021).
\newblock {Exact solutions and equi-dosing regimen regions for multi-dose
  pharmacokinetics models with transit compartments}.
\newblock {\em J. Pharmacokinet. Pharmacodyn.}, 48(1):99--131.

\bibitem[Hogan et~al., 2013]{hogan2013clonally}
Hogan, T., Shuvaev, A., Commenges, D., Yates, A., Callard, R., Thiebaut, R.,
  and Seddon, B. (2013).
\newblock Clonally diverse T cell homeostasis is maintained by a common program
  of cell-cycle control.
\newblock {\em J. Immunol}, 190(8):3985--3993.

\bibitem[Holder and Beauchemin, 2011]{Holder2011}
Holder, B.~P. and Beauchemin, C.~A. (2011).
\newblock {Exploring the effect of biological delays in kinetic models of
  influenza within a host or cell culture}.
\newblock {\em BMC Public Health}, 11(Suppl 1):S10.

\bibitem[Hu et~al., 2018]{Hu2018}
Hu, S., Dunlavey, M., Guzy, S., and Teuscher, N. (2018).
\newblock {A distributed delay approach for modeling delayed outcomes in
  pharmacokinetics and pharmacodynamics studies}.
\newblock {\em J. Pharmacokinet. Pharmacodyn.}, 45(2):1--24.

\bibitem[Hurtado and Richards, 2020]{Hurtado2020}
Hurtado, P. and Richards, C. (2020).
\newblock {Time Is Of The Essence: Incorporating Phase-Type Distributed Delays
  And Dwell Times Into ODE Models}.
\newblock {\em Math. Appl. Sci. Eng.}, 1(4):410--422.

\bibitem[Hurtado and Kirosingh, 2019]{Hurtado2019}
Hurtado, P.~J. and Kirosingh, A.~S. (2019).
\newblock {Generalizations of the `Linear Chain Trick': incorporating more
  flexible dwell time distributions into mean field ODE models}.
\newblock {\em J. Math. Biol.}, 79(5):1831--1883.

\bibitem[Kinjyo et~al., 2015]{kinjyo2015real}
Kinjyo, I., Qin, J., Tan, S.-Y., Wellard, C.~J., Mrass, P., Ritchie, W., Doi,
  A., Cavanagh, L.~L., Tomura, M., Sakaue-Sawano, A., et~al. (2015).
\newblock Real-time tracking of cell cycle progression during CD8+ effector and
  memory T-cell differentiation.
\newblock {\em Nat. Commun.}, 6(1):6301.

\bibitem[Koch et~al., 2014]{Koch2014}
Koch, G., Krzyzanski, W., P{\'{e}}rez-Ruixo, J.~J., and Schropp, J. (2014).
\newblock {Modeling of delays in PKPD: classical approaches and a tutorial for
  delay differential equations}.
\newblock {\em J. Pharmacokinet. Pharmacodyn.}, 41(4):291--318.

\bibitem[Koch and Schropp, 2015]{Koch2015}
Koch, G. and Schropp, J. (2015).
\newblock {Distributed transit compartments for arbitrary lifespan
  distributions in aging populations}.
\newblock {\em J. Theor. Biol.}, 380:550--558.

\bibitem[Krzyzanski, 2011]{Krzyzanski2011}
Krzyzanski, W. (2011).
\newblock {Interpretation of transit compartments pharmacodynamic models as
  lifespan based indirect response models.}
\newblock {\em J. Pharmacokinet. Pharmacodyn.}, 38(2):179--204.

\bibitem[Krzyzanski, 2019]{Krzyzanski2019}
Krzyzanski, W. (2019).
\newblock {Ordinary differential equation approximation of gamma distributed
  delay model}.
\newblock {\em J. Pharmacokinet. Pharmacodyn.}, 46(1):53--63.

\bibitem[Krzyzanski et~al., 2018]{Krzyzanski2018}
Krzyzanski, W., Hu, S., and Dunlavey, M. (2018).
\newblock {Evaluation of performance of distributed delay model for
  chemotherapy-induced myelosuppression}.
\newblock {\em J. Pharmacokinet. Pharmacodyn.}, 45(2):329--337.

\bibitem[Liao et~al., 2020]{Liao2020}
Liao, L.~E., Carruthers, J., Smither, S.~J., Weller, S.~A., Williamson, D.,
  Laws, T.~R., Garc{\'{i}}a-Dorival, I., Hiscox, J., Holder, B.~P., Beauchemin,
  C. A.~A., Perelson, A.~S., L{\'{o}}pez-Garc{\'{i}}a, M., Lythe, G., Barr,
  J.~N., and Molina-Par{\'{i}}s, C. (2020).
\newblock {Quantification of Ebola virus replication kinetics in vitro}.
\newblock {\em PLOS Comput. Biol.}, 16(11):e1008375.

\bibitem[Lyons and Parish, 1994]{lyons1994determination}
Lyons, A.~B. and Parish, C.~R. (1994).
\newblock Determination of lymphocyte division by flow cytometry.
\newblock {\em J. Immunol. Methods}, 171(1):131--137.

\bibitem[MacDonald, 1978]{MacDonald1978}
MacDonald, N. (1978).
\newblock {Time Lags in Biological Models}.

\bibitem[Mackey, 1978]{Mackey1978}
Mackey, M.~C. (1978).
\newblock {Unified hypothesis for the origin of aplastic anemia and periodic
  hematopoiesis}.
\newblock {\em Blood}, 51(5):941--956.

\bibitem[Mackey and Glass, 1977]{Mackey1977}
Mackey, M.~C. and Glass, L. (1977).
\newblock {Oscillation and chaos in physiological control systems}.
\newblock {\em Science (80-. ).}, 197(4300):287--289.

\bibitem[MATLAB, 2017]{Matlab2017}
MATLAB (2017).
\newblock {R2017a}.

\bibitem[Mazzocco et~al., 2017]{mazzocco2017estimates}
Mazzocco, P., Bernard, S., and Pujo-Menjouet, L. (2017).
\newblock Estimates and impact of lymphocyte division parameters from cfse data
  using mathematical modelling.
\newblock {\em PLOS One}, 12(6):e0179768.

\bibitem[Perelson, 2002]{Perelson2002}
Perelson, A.~S. (2002).
\newblock {Modelling viral and immune system dynamics}.
\newblock {\em Nat. Rev. Immunol.}, 2(1):28--36.

\bibitem[Perelson et~al., 1996]{Perelson1996}
Perelson, A.~S., Neumann, A.~U., Markowitz, M., Leonard, J.~M., and Ho, D.~D.
  (1996).
\newblock {HIV-1 dynamics in vivo: Virion clearance rate, infected cell
  life-span, and viral generation time}.
\newblock {\em Science.}, 271(5255):1582--1586.

\bibitem[P{\'{e}}rez-Ruixo et~al., 2012]{Perez-Ruixo2012}
P{\'{e}}rez-Ruixo, C., Valenzuela, B., {Fern{\'{a}}ndez Teruel}, C.,
  Gonz{\'{a}}lez-Sales, M., Miguel-Lillo, B., Soto-Matos, A., and
  P{\'{e}}rez-Ruixo, J.~J. (2012).
\newblock {Population pharmacokinetics of PM00104 (Zalypsis{\textregistered})
  in cancer patients}.
\newblock {\em Cancer Chemother. Pharmacol.}, 69(1):15--24.

\bibitem[Porthiyas et~al., 2024]{Porthiyas2024}
Porthiyas, J., Nussey, D., Beauchemin, C. A.~A., Warren, D.~C., Quirouette, C.,
  and Wilkie, K.~P. (2024).
\newblock {Practical parameter identifiability and handling of censored data
  with Bayesian inference in mathematical tumour models}.
\newblock {\em npj Syst. Biol. Appl.}, 10(1):89.

\bibitem[Quartino et~al., 2014]{Quartino2014}
Quartino, A.~L., Karlsson, M.~O., Lindman, H., and Friberg, L.~E. (2014).
\newblock {Characterization of endogenous G-CSF and the inverse correlation to
  chemotherapy-induced neutropenia in patients with breast cancer using
  population modeling}.
\newblock {\em Pharm. Res.}, 31(12):3390--3403.

\bibitem[Quirouette et~al., 2023]{Quirouette2023}
Quirouette, C., Cresta, D., Li, J., Wilkie, K.~P., Liang, H., and Beauchemin,
  C. A.~A. (2023).
\newblock {The effect of random virus failure following cell entry on infection
  outcome and the success of antiviral therapy}.
\newblock {\em Sci. Rep.}, 13(1):17243.

\bibitem[Raue et~al., 2009]{Raue2009}
Raue, A., Kreutz, C., Maiwald, T., Bachmann, J., Schilling, M.,
  Klingm{\"{u}}ller, U., and Timmer, J. (2009).
\newblock {Structural and practical identifiability analysis of partially
  observed dynamical models by exploiting the profile likelihood}.
\newblock {\em Bioinformatics}, 25(15):1923--1929.

\bibitem[Ribeiro et~al., 2010]{Ribeiro2010}
Ribeiro, R.~M., Qin, L., Chavez, L.~L., Li, D., Self, S.~G., and Perelson,
  A.~S. (2010).
\newblock {Estimation of the Initial Viral Growth Rate and Basic Reproductive
  Number during Acute HIV-1 Infection}.
\newblock {\em J. Virol.}, 84(12):6096--6102.

\bibitem[Roskos et~al., 2006]{Roskos2006}
Roskos, L.~K., Lum, P., Lockbaum, P., Schwab, G., and Yang, B.-B. (2006).
\newblock {Pharmacokinetic/pharmacodynamic modeling of pegfilgrastim in healthy
  subjects}.
\newblock {\em J. Clin. Pharmacol.}, 46(7):747--757.

\bibitem[Rubinow and Lebowitz, 1975]{Rubinow1975}
Rubinow, S. and Lebowitz, J. (1975).
\newblock {A mathematical model of neutrophil production and control in normal
  man}.
\newblock {\em J. Math. Biol.}, 225:187--225.

\bibitem[Scarabel et~al., 2026]{Scarabel2025}
Scarabel, F., Coldwell, H., and Cassidy, T. (2026).
\newblock {Genuine and spurious bistability in a simple epidemic model with
  waning immunity}.
\newblock {\em Math. Biosci.}, 396:109671.

\bibitem[{Simulations Plus}, 2023]{Monolix2020}
{Simulations Plus} (2023).
\newblock {Monolix}.

\bibitem[Soto et~al., 2020]{Soto2020}
Soto, E., Banfield, C., Gupta, P., and Peterson, M.~C. (2020).
\newblock {Kinetic-Pharmacodynamic Model of Platelet Time Course in Patients
  With Moderate-to-Severe Atopic Dermatitis Treated With Oral Janus Kinase 1
  Inhibitor Abrocitinib}.
\newblock {\em CPT Pharmacometrics Syst. Pharmacol.}, 9(10):553--560.

\bibitem[Vehtari et~al., 2017]{Vehtari2017}
Vehtari, A., Gelman, A., and Gabry, J. (2017).
\newblock {Practical Bayesian model evaluation using leave-one-out
  cross-validation and WAIC}.
\newblock {\em Stat. Comput.}, 27(5):1413--1432.

\bibitem[Vogel, 1961]{Vogel1961}
Vogel, T. (1961).
\newblock {Syst{\`{e}}mes D{\'{e}}ferlants, Syst{\`{e}}mes
  H{\'{e}}r{\'{e}}ditaires, Syst{\`{e}}mes Dynamiques}.
\newblock {\em Proc. Int. Symp. Nonlinear Vib.}, pages 123--130.

\bibitem[von Schulthess and Mazer, 1982]{VonSchulthess1982}
von Schulthess, G. and Mazer, N. (1982).
\newblock {Cyclic neutropenia (CN): A clue to the control of granulopoiesis}.
\newblock {\em Blood}, 59:27--37.

\bibitem[Wearing et~al., 2005]{Wearing2005}
Wearing, H.~J., Rohani, P., and Keeling, M.~J. (2005).
\newblock {Appropriate Models for the Management of Infectious Diseases}.
\newblock {\em PLoS Med.}, 2(7):e174.

\bibitem[Yates et~al., 2017]{Yates2017}
Yates, C.~A., Ford, M.~J., and Mort, R.~L. (2017).
\newblock {A Multi-stage Representation of Cell Proliferation as a Markov
  Process}.
\newblock {\em Bull. Math. Biol.}, 79(12):2905--2928.

\end{thebibliography}

\setcounter{equation}{0}
\setcounter{figure}{0}
\setcounter{table}{0}
\makeatletter
\renewcommand{\theequation}{S\arabic{equation}}
\renewcommand{\thefigure}{S\arabic{figure}}
\renewcommand{\thetable}{S\arabic{table}}
\renewcommand{\thetheorem}{S\arabic{theorem}}
\setlength\parindent{2em}

\section*{Supplemental Information}


\begin{figure}[h!]
\centering
\includegraphics[width=0.8\textwidth]{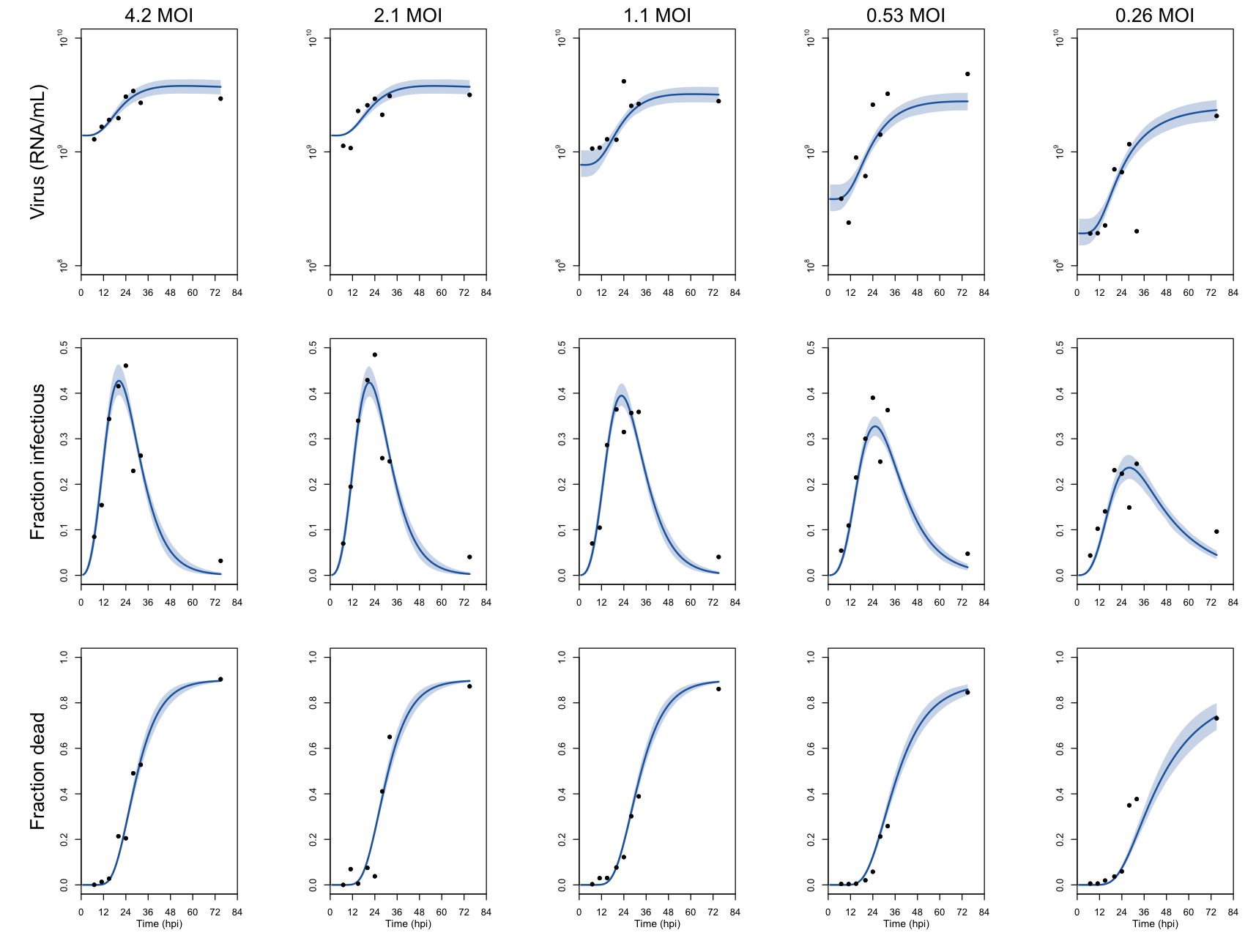}
\caption{Posterior-predictive dynamics across MOI doses for the \citet{Beauchemin2017} SHIV data. The experimental data is plotted as filled circles across the five MOIs, while the solid lines are posterior mean trajectories and shaded bands represent the 90\% credible intervals for the dynamics.
}
\label{Fig:stan_dynamics}
\end{figure} 

\end{document}